\documentclass[a4paper,11pt]{article}
\pdfoutput=1       
\usepackage{a4wide}
\usepackage{ascii}
\usepackage[T1]{fontenc}
\usepackage{amsmath,amssymb,amsthm,mathtools,array,booktabs}
\usepackage{stmaryrd}
\usepackage{graphicx,hyperref}
\usepackage{xcolor}
\usepackage{cancel}
\usepackage{tikz}
\usetikzlibrary{cd}
\usepackage{amscd}
\usepackage{latexsym,cite}
\usepackage{subcaption}
\usepackage[boxsize=0.8em]{ytableau}
\usepackage{pictexwd,dcpic}
\usepackage{youngtab}

\usepackage{scalerel}   
\newcommand{\bigboxtimes}{%
  \mathop{\scalerel*{\boxtimes}{\bigotimes}}\displaylimits
}

\Yboxdim{5pt}

\input xy
\xyoption{all}

\hypersetup{
  unicode,
  bookmarksnumbered,
  linktoc = all,
  pdfborderstyle = {/S/U/W 0.5}
}

\newcommand{\bbC}{\mathbb{C}}

\newcommand{\bbG}{\mathbb{G}}

\newcommand{\bbP}{\mathbb{P}}

\newcommand{\bbZ}{\mathbb{Z}}

\newcommand{\calB}{\mathcal{B}}
\newcommand{\calC}{\mathcal{C}}

\newcommand{\calE}{\mathcal{E}}

\newcommand{\calL}{\mathcal{L}}
\newcommand{\calM}{\mathcal{M}}
\newcommand{\calN}{\mathcal{N}}
\newcommand{\calO}{\mathcal{O}}

\newcommand{\calR}{\mathcal{R}}
\newcommand{\calS}{\mathcal{S}}

\newcommand{\calW}{\mathcal{W}}

\DeclareMathOperator{\Ch}{ch}

\DeclareMathOperator{\Cone}{Cone}

\DeclareMathOperator{\Id}{Id}

\DeclareMathOperator{\rank}{rank}

\DeclareMathOperator{\Sym}{Sym}
\DeclareMathOperator{\Td}{Td}

\DeclareMathOperator{\Tr}{Tr}

\newtheorem{thm}{\color{blue} Theorem} \numberwithin{thm}{section}

  \numberwithin{rmk}{subsection}

  \numberwithin{eg}{subsection}

 \numberwithin{conj}{section}

\newtheorem{prop}{\color{blue} Proposition} \numberwithin{prop}{section}

 \numberwithin{lem}{section}

 \numberwithin{cor}{section}

 \numberwithin{defi}{section}

\begin{document}
\thispagestyle{empty}
\begin{flushright}

\end{flushright}
\vspace{1cm}
\begin{center}
{\LARGE\bf GLSM monodromy on quantum period lattice of Calabi-Yau fourfold flops} 
\end{center}
\vspace{8mm}
\begin{center}
Ban Lin\footnote{{\tt banlin@kias.re.kr}}$^{\dagger}$
\end{center}
\vspace{6mm}
\begin{center}
$^\dagger$Korea Institute for Advanced Study, Seoul, Republic of Korea, 02455
\end{center}
\vspace{15mm}

\begin{abstract}
\noindent

We establish an integral basis for the B-brane central charges of certain Calabi–Yau fourfold flops and derive exact expressions for their monodromies using the grade restriction rule and window categories of the associated gauged linear sigma models. The monodromy is interpreted as an EZ twist associated with the contraction of an exceptional surface onto a curve of conifold singularities. We further illustrate a decomposition of the EZ twist into spherical twists for families of splitting configurations of the sixtic Calabi–Yau fourfold and a complete intersection in grassmannian.

\end{abstract}
\newpage
\setcounter{tocdepth}{3}
\tableofcontents
\setcounter{footnote}{0}

\section{Introduction}

Calabi--Yau fourfolds arise naturally as compactification spaces of
M-theory and F-theory, and also define two-dimensional
$\mathcal N=(2,2)$ theories by compactification of type IIA string
theory. Their topology determines the spectrum of effective fields,
the lattice of admissible background fluxes, and the charges of wrapped
branes. It is therefore important to understand how these data change
under transitions between different birational or deformation phases.\cite{Green:1988wa,Green:1996dd} In the presence of
background flux, fourfold conifold transitions also relate different
Coulomb and Higgs phases of the effective three-dimensional $\calN=2$ theories
\cite{Intriligator:2012ue}.

One of the geometric transition is provided by the flop transition in a splitting configuration in the product of grassmannians $\bbG=\prod_\alpha G(k_\alpha,m_\alpha)$:
\[
 X=
 \left[
 \begin{array}{c|cccccc}
 \mathbb P^{m-1} & 1 & \cdots & 1 & 0 & \cdots & 0\\
 \bbG               &  n_1 & \cdots &  n_m
                 &  n_{m+1} & \cdots &  n_K
 \end{array}
 \right].
\]
Projection along the $\mathbb P^{m-1}$ direction contracts
$X$ to a determinantal Calabi--Yau fourfold $X^\sharp$.
Under genericity assumptions, its singular locus is a curve
$\Sigma$ of ordinary conifold singularities, and the exceptional locus
of the small resolution is a surface fibered over $\Sigma$ with generic
fiber $\mathbb P^1$. The second small resolution $\widetilde X$ is
obtained by flopping this family of rational curves. This differs from
the familiar Calabi--Yau threefold flop, where the exceptional locus is
generically a finite collection of curves. A deformation of
$X^\sharp$, on the other hand, gives the unsplit fourfold
\[
 X^\flat=
 \left[
 \begin{array}{c|cccc}
 \bbG &  n & n_{m+1}
   & \cdots &  n_K
 \end{array}
 \right], \quad n=n_1+\cdots+n_m.
\]
so that the resolution and deformation phases form a fourfold conifold
transition. The topology change is controlled by the geometry of
$\Sigma$ and by the ruled exceptional surface
\cite{Brunner:1996bu,Intriligator:2012ue}.\\

The flop geometry $X$ and $\widetilde{X}$ can be realized as different phases $X_{\zeta_\pm}$ in a single gauged linear sigma model (GLSM) known as the abelian PAX model \cite{Jockers:2012zr}. We consider a four dimensional CY complete intersection (CYCI):
\begin{equation}
X_{\zeta_{+}}\subset \mathbb{P}^{m-1}\times \bbG,\qquad \bbG:=G(k_{1},m_{1})\times\cdots\times G(k_{r},m_{r})
\end{equation}
defined in \eqref{Xplus}, which can be realized as the geometric phase of a gauge linear sigma model (GLSM), described in sect. \ref{sec:setting}, with gauge group $G=U(1)_{0}\times U(k_{1})\times\ldots U(k_{r})$. Such GLSM have $r+1$ FI-theta parameters, denoted as
\begin{equation}
t_{\alpha}=\zeta_{\alpha}-i\theta_{\alpha},\qquad\alpha=0,\ldots, r
\end{equation}
they are coordinates in the stringy K\"ahler moduli $\calM_{K}$ of such a model. We will focus on two phases in $\calM_{K}$, namely:
\begin{equation}
\zeta_{\pm}:=\{\zeta_{0}\gg \pm1,\zeta_{1}\gg1,\ldots\zeta_{r}\gg1\}
\end{equation}
then, the $\zeta_{+}$ RG flows to a nonlinear sigma model with target space the aforementioned CICY $X=X_{\zeta_{+}}$. We use the grade restriction rule \cite{Herbst:2008jq,Hori:2013ika} to find an appropriate set of generators for $D^b\mathrm{Coh}(X_{\zeta_{+}})$. More precisely, they correspond to an integral basis of B-brane charges as \cite{Gerhardus:2016iot}
\begin{eqnarray}
    \vec\Pi=\left( Z_{X},\ Z_{D_\alpha},\ Z_{S_\Gamma},\ Z_{\widetilde{C_\alpha}},\ Z_P \right),\quad \alpha=0,1,\cdots,r,
\end{eqnarray}
corresponding to the holomorphic cycles of $D8,D6,D4,D2,D0$ branes on $X$. We also establish a dual surface basis $\calO_{\widetilde{S_\Gamma}}$ with generators homotopic to $\bbP^2$ and $\bbP^1\times\bbP^1$ in the class of exceptional surfaces. The monodromy computation in section \ref{sec:monodromy} can be summarized as in theorem \ref{thm:monodromy}: The window shift monodromy action $M$ on the integral charge lattice $\vec\Pi$ acts as
    \begin{eqnarray}
        \vec\Pi\cdot M=\vec\Pi\cdot T\cdot L_{K_Y}
    \end{eqnarray}
for $L_{K_Y}(\calB)=\calB\otimes K_Y$ is the action of tensoring the canonical bundle of $Y$ (see \eqref{eqn:L}), in where $X^\flat$ is the Calabi-Yau hypersurface:
\begin{eqnarray}
    Y=\left[
 \begin{array}{c|cccc}
 \bbG  & n_{m+1}
   & \cdots &  n_K
 \end{array}
 \right].
\end{eqnarray}
Meanwhile, the action $T$ acts on divisor class $\calO_{D_0}$ and surface classes (labeled by $\Gamma=0\alpha$) $\calO_{S_{0\alpha}}$ for $\alpha=0,1,\cdots,r$, as (see \eqref{eqn:T})
\begin{eqnarray}
   \begin{pmatrix}
        Z_{D_0^{\;}}
        \\ 
        Z_{S_{00}^{\;}}
        \\
        Z_{S_{0\alpha}^{\;}} 
    \end{pmatrix}^t\cdot T=
    \begin{pmatrix}
        Z_{D_0}-N_5\cdot Z_{\widetilde{S_{00}}}-N_4^{(\beta)}Z_{\widetilde{S_{0\beta}}}-(N_4/2) Z_{\widetilde{C_0}}
        \\
        Z_{S_{00}}-N_5Z_{\widetilde{C_0}}
        \\
        Z_{S_{0\alpha}}-N_4^{(\alpha)}Z_{\widetilde{C_0}}
    \end{pmatrix}^t
\end{eqnarray}
which is an EZ-twist of the collaping of exceptional surface in the class $[S]=N_5[\widetilde{S_{00}}]+\sum_\alpha N_4^{(\alpha)}[\widetilde{S_{0\alpha}}]$ to a genus $1-g=N_4/2$ curve with a generic $\bbP^1$-fiber in $[\widetilde{C_0}]$ \cite{Aspinwall:2001zq,Horja:2001cp}. 

Such a monodromy result is conjecturally related to the topology of Kahler moduli space. Using the similar argument in \cite{Lin:2024fpz,Lin:2026icu}, we propose some braid group relations on some examples that $M$ should satisfies. In particular, we find the following relations:
\begin{itemize}
    \item For $X$ as splitting configurations of $X^\flat=\bbP^5[6]$, $L=-\otimes\calO_{X}(1)$, $L_{K_Y}\equiv L^{-6}$, and $T_{\calO_X}$ the spherical twist of structure sheaf $\calO_X$ \cite{Seidel:2000ia}, one has
    \begin{eqnarray}
        M=T\cdot L_{K_Y}\equiv (T_{\calO_X}\cdot L)^6 L^{-6}.
    \end{eqnarray}

    \item For $X$ as splitting configurations of $X^\flat=G(2,5)[4,1]$ of the degree four hyperplane, $L=-\otimes\calO_{X}(1)$, $L_{K_Y}\equiv L^{-4}$, and $T_{\calO_X}$ and $T_{\calS_X}$ are the spherical twist of structure sheaf $\calO_X$ and tautological bundle $\calS_X=\calS|_X$, one has
    \begin{eqnarray}
        M=T\cdot L_{K_Y}\equiv (T_{\calS_X}\cdot T_{\calO_X}\cdot L)^4 L^{-4}.
    \end{eqnarray}

    In addition, for the splitting of the degree one hyperplane in $X^\flat$, we discover the following result instead:
    \begin{eqnarray}
        T^4\equiv (T_{\calS_X}\cdot T_{\calO_X}\cdot L)^4.
    \end{eqnarray}
\end{itemize}

The paper is organized as follows. In section \ref{sec:setting} we first review the gauged linear sigma model and GLSM brane for splitting Calabi-Yau varieties, then the window shift monodromy from grade restriction rule and its decomposition proposal in Kahler moduli space. In section \ref{sec:period}, we discuss in detail the classical part of A-period lattice for Calabi-Yau fourfolds. In addition, we propose another set of generators $\calO_{\widetilde{S}}$ in $H^4(X,\bbC)$ that are dual to surface classes $\calO_S$. A list of the topological data for splitting Calabi-Yau fourfolds is also conducted. In section \ref{sec:monodromy}, using the partial results theorem \ref{thm:monodromyCY3} from Calabi-Yau threefolds, the window shift monodromy action $M$ is computed on the A-period lattice as in theorem \ref{thm:monodromy}. Finally, we illustrate by examples the decomposition of monodromy $M$ into the braid group action of spherical twists, of two splitting families over the sixtic fourfold $\bbP^5[6]$ and the grassmannian complete intersection $G(2,5)[4,1]$.

\section*{Acknowledgements}

We thank Cyril Closset and Ed Segal for enlightening discussions, and the hospitality and support from the Mainz Institute for Theoretical Physics of the Cluster of Excellence PRISMA$^+$ (Project ID 390831469). The author especially thanks Mauricio Romo for collaboration on a related project \cite{Lin:2026icu} and Zekai Yu for helpful interaction at the early stage of this work. Ban Lin is supported by the KIAS Individual Grant (PG100701) at Korea Institute for Advanced Study.

\section{GLSM Settings} \label{sec:setting}

In this section, we review some necessary background on the gauged linear sigma model that realizes the flop in splitting Calabi-Yau varieties \cite{Lin:2026icu}. In particular, it is the abelian PAX model for determinantal Calabi-Yau discussed in \cite{Jockers:2012zr}. A gauged linear sigma model consists of the following data:
\begin{itemize}
    \item A gauge group 
    \begin{equation}
        G=U(1)\times U(k_{1})\times\cdots\times U(k_{r}).
    \end{equation}

    \item A complex vector space $V$ with a representation $\rho:G\rightarrow GL(V)$ and a representation $R:\bbC^*\rightarrow GL(V)$ of (vector) $U(1)_R$ symmetry. The Calabi-Yau condition for the GLSM to be anomaly-free is that $\rho$ factors through $SL(V)$.

    \item A superpotential in gauge-invariant polynomial $W\in\Sym(V^\vee)^G$ with $U(1)_R$ charge 2, which will be given momentarily.

    \item A twisted superpotential characterized by the FI-$\theta$ parameters $t=\zeta-i\theta$ for \cite{Hori:2019vkm}
     \begin{eqnarray}
  t\in \left(\frac{\mathfrak{t}^{\vee}_{\mathbb{C}}}{2\pi i
\mathrm{P}}\right)^{W_{G}}\cong\frac{\mathfrak{z}^{\vee}_{\mathbb{C}}}{2\pi i
\mathrm{P}^{W_{G}}}.
  \end{eqnarray}
  Here $\mathrm{P}$ denotes the weight lattice, $W_{G}$ the Weyl subgroup of
$G$, $\mathfrak{t}$ the Cartan subalgebra of $\mathfrak{g}=\mathrm{Lie}(G)$ and
$\mathfrak{z}=\mathrm{Lie}(Z(G))$. In the present case we can choose a basis where the FI-theta parameters for the $U(1)$ and $U(k_{\alpha})$ subgroups of $G$, can be denoted $t_0$ and $t_\alpha$, $\alpha=1,\ldots,r$, respectively.
    
\end{itemize}

In abelian PAX models, we fix the representation $\rho$ as the following
\begin{eqnarray}
      V\equiv \mathbb{C}(1)^{\oplus m}\oplus\mathbf{k}_{1}^{\oplus m_{1}}\oplus\cdots \oplus \mathbf{k}_{r}^{\oplus m_{r}}\bigoplus_{I=1}^{m}(\mathbb{C}(-1)\otimes \mathcal{R}_{I})\bigoplus_{I=m+1}^{K}\mathcal{R}_{I} \label{eqn:chiralfields}
    \end{eqnarray}
for $ m_{\alpha}\in \mathbb{Z}_{\geq k_{\alpha}},\ K\in\mathbb{N}_{\geq m}$. And $\mathcal{R}_{I}$ denotes the rank one representation:\footnote{We assume that $\calR_I$ has rank one here only for simplicity. In the later discussion, $\calR_I$ can carry other representations of $G$.}
\begin{eqnarray}
\mathcal{R}_{I}:=\mathrm{det}^{-n^{(1)}_{I}}\otimes\cdots \otimes \mathrm{det}^{-n^{(r)}_{I}},
\end{eqnarray}
with $n^{(\alpha)}_{I}\in\mathbb{N}$ and $\mathrm{det}^{-n^{(\alpha)}_{I}}$ denoting the $-n^{(\alpha)}_{I}$th power of the determinant representation of $U(k_{\alpha})\subset G$. The Calabi-Yau condition is thus
\begin{eqnarray}
\sum_{I=1}^{K}n^{(\alpha)}_{I}=m_{\alpha},\qquad \alpha=1,\ldots,r.
\end{eqnarray}
We denote the coordinates in $V$ as
\begin{eqnarray}
x^{(\alpha)}&&\text{ \ \ coordinates on \ \ }\mathbf{k}_{\alpha}^{\oplus m_{\alpha}},\qquad \alpha=1,\ldots,r\nonumber\\
p_{I}&&\text{ \ \ coordinates on \ \ }\mathcal{R}_{I},\qquad I=1,\ldots, K\nonumber\\
y &&\text{ \ \ coordinates on \ \ } \mathbb{C}(1)^{\oplus m},
\end{eqnarray}
Then we can define the $G$-invariant superpotential $W$ to specify the matter interaction:\footnote{The determinantal superpotential $W=\Tr p A(x) y$ gives the name of "PAX model", while another type of such superpotential $W=\Tr p(A-xy)$ gives the "PAXY model" \cite{Jockers:2012zr}.}
\begin{equation}\label{superpot}
W:=\sum_{I=1}^K p_{I}F_{I}(x,y)=\sum_{I=1}^{m}p_IF_I^J(x)y_J+\sum_{J=m+1}^K p_JF_J(x),
\end{equation}
where
\begin{eqnarray}
F_{I}(x,y)&&\text{ \ \ homogeneous of degree \ \ }(1, n^{(1)}_{I},\ldots,n^{(r)}_{I}),\qquad I=1,\ldots,m\nonumber\\
F_{I}(x)&&\text{ \ \ homogeneous of degree \ \ }(0, n^{(1)}_{I},\ldots,n^{(r)}_{I}),\qquad I=m+1,\ldots,K.
\end{eqnarray}\\

In summary, the matter content of GLSM is indicated in the following table:
\begin{equation}
    \begin{array}{c|cccccc}
         & x^{(1)} & \cdots & x^{(r)} & y_{1,\cdots,m} & p_{I=1,\cdots,m} & p_{I=m+1,\cdots,K}
         \\\hline
       U(1)  & 0 & \cdots & 0 & 1 & -1 & 0 
       \\
       U(k_1) & \mathbf{k}^{m_1}_1 & \cdots & 0 & 0 & \mathrm{det}^{-n^{(1)}_I} & \mathrm{det}^{-n^{(1)}_I}
       \\
       \vdots &  & \ddots & & \vdots & \vdots & \vdots
       \\
       U(k_r) & 0 & \cdots & \mathbf{k}_r^{m_r} & 0 & \mathrm{det}^{-n^{(r)}_I} & \mathrm{det}^{-n^{(r)}_I}
       \\\hline
       U(1)_R & \varepsilon_1 & \cdots & \varepsilon_r & \varepsilon_0 & \epsilon_{I} & \widetilde \epsilon_{I}
    \end{array}
\end{equation}
where the R-charges are denoted $\varepsilon_0,\varepsilon_\alpha,\epsilon_I,\widetilde \epsilon_I\in[0,2)$, satisfying
\begin{equation} 
    \begin{array}{ll}
     \epsilon_I=2-\varepsilon_0-2\sum_{\alpha=1}^rn_I^{(\alpha)}\varepsilon_\alpha,   &  \quad I=1,\cdots,m,
         \\
    \widetilde \epsilon_I= 2-2\sum_{\alpha=1}^rn_I^{(\alpha)}\varepsilon_\alpha,     &  \quad I=m+1,\cdots,K.
    \end{array}
   \label{eqn:Rcharge}
\end{equation}
In general, the weights of $R$ cannot be fixed in a GLSM, since it is subjected to ambiguities. At an IR fixed point, it is expected that the weights of $R$ are fixed, under RG flow, to values in the interval $(0,2)$ \cite{Lerche:1989uy}. These values can be different, for different phases (or chambers) in the stringy K\"ahler moduli $\mathcal{M}_{K}$ (defined below).

\subsection{Geometric phases}

This GLSM, by construction have a classical Higgs phase on the regime $\zeta_{0},\ldots,\zeta_{k}\gg 1$, given by
\begin{equation}
X_{\zeta_{+}}:=\mu^{-1}(\zeta)/G\cap \{\mathrm{Crit}W\}
\end{equation}
where $\mu:V\rightarrow \mathfrak{g}^{\vee}$ denotes the moment map, associated to $\rho$, on the
vector space $V$ (whose coordinates are $(y,x,p)$). For generic polynomials $F_{I}(x,y)$, $X_{\zeta_{+}}$ is given by the smooth complete intersection
\begin{equation}\label{Xplus}
X_{\zeta_{+}}=\bigcap_{I=1}^{K}\{F_{I}(x,y)=0\}\subset \mathbb{P}^{m-1}\times G(k_{1},m_{1})\times\cdots\times G(k_{r},m_{r})
\end{equation}
The central charge $\hat{c}$ of this GLSM coincide with $d=\mathrm{dim}X_{\zeta_{+}}$:
\begin{equation}
\hat{c}=d=m-1+\sum_{\alpha=1}^{r}k_{\alpha}(m_{\alpha}-k_{\alpha})-K
\end{equation}
We will use the following notation for the complete intersection CY \cite{Green:1986ck,Candelas:1987kf}(CICY) $X_{\zeta_{+}}$:
\begin{equation}
\begin{aligned}
    X_{\zeta_{+}}=\left[ \begin{array}{c|cccccc}
      \bbP^{m-1}   &  1 & \cdots & 1 & 0 & \cdots & 0
    \\
       \bbG  &  {n}_1 & \cdots & {n}_{m} &  {n}_{m+1} & \cdots &  {n}_K
    \end{array}\right],
    \label{eqn: CICYI}
\end{aligned}
\end{equation}
where $\bbG:=G(k_{1},m_{1})\times\cdots\times G(k_{r},m_{r})$ and $n_I=(n_I^{(1)},\cdots,n_I^{(r)})$ . For the abelian case: $k_{\alpha}=1$ for all $\alpha$, these configurations have been studied in \cite{Candelas:1987kf,Green:1988wa,Brodie:2021toe} as examples of global flops in CY3 and termed \emph{splitting configurations}. The phase $-\zeta_{0},\zeta_{1},\ldots,\zeta_{k}\gg 1$ will also be relevant in our analysis, so we proceed to describe it here. The analysis goes likewise. It is also a pure Higgs phase described by a sigma model with target space $X_{\zeta_{-}}$ which is a complete intersection:
\begin{eqnarray}
X_{\zeta_{-}}&=&\bigcap_{I=1}^{m}\{F'_{I}(x,p)=0\}\bigcap_{I=m+1}^{K}\{F_{I}(x)=0\}\subset\mathbb{P}_{-} \nonumber\\
\mathbb{P}_{-}&:=&\mathbb{P}\left(\bigoplus_{I=1}^{m}\mathcal{O}(-n^{(1)}_{I},\ldots,-n^{(r)}_{I})\right)\rightarrow G(k_{1},m_{1})\times\cdots\times G(k_{r},m_{r})
\end{eqnarray}
where $\mathcal{O}(-n^{(\alpha)}_{I})$ denotes the line bundle $(\mathrm{det}\calS_{\alpha})^{\otimes n^{(\alpha)}_{I}}\rightarrow G(k_{\alpha},m_{\alpha})$ with $\calS_{\alpha}$ the tautological bundle of rank $k_{\alpha}$ and,
\begin{eqnarray}
F'_{I}(x,p):=\frac{\partial}{\partial y_{I}} \sum_{J=1}^{m}p_{J}F_{J}(x,y)
\end{eqnarray}

\textbf{Comment on vector R-charges on $\zeta_{\pm}$ phases:} From \eqref{eqn:Rcharge}, the vector R-charge for matter fields in the phases $X_{\zeta_{\pm}}$ can be assigned as
\begin{equation}
    \begin{array}{cl}
        (\varepsilon_\alpha,\ \varepsilon_0,\ \epsilon_I,\ \widetilde\epsilon_I)=(0,0,2,2) &  \text{for } \zeta_+ \text{ \ phase}
         \\
        (\varepsilon_\alpha,\ \varepsilon_0,\ \epsilon_I,\ \widetilde\epsilon_I)=(0,2,0,2) & \text{for }\zeta_- \text{ \ phase}
    \end{array}\label{eqn:RInt}
\end{equation}
These R-charges are not strictly inside the interval $(0,2)$, however, in this limit, physical correlators such as the hemisphere partition function are well defined, so we will work with R-charges \eqref{eqn:RInt}, whenever we need an explicit assignment.

\subsection{Effective twisted potential}

The effective twisted potential is the potential for the scalar fields $\sigma$ in the vector multiplet \cite{Witten:1993yc,Morrison:1994fr}. In a generic region of the Coulomb branch, i.e. where the eigenvalues of the VEV $\langle\sigma\rangle\in\mathfrak{t}_{\mathbb{C}}:=\mathrm{Lie}(T_{G})\otimes \mathbb{C}$ are distinct and their magnitude is large (compared with the energy scale). Upon integration of the charged chirals, the effective twisted potential is given by \cite{Hori:2019vkm,Morrison:1994fr}
\begin{equation}\label{genericWeff}
\widetilde{W}_{\mathrm{eff}}(\sigma)=-t(\sigma)+\pi i\sum_{\alpha>0} \alpha(\sigma)-\sum_{\sf{a}}Q_{\sf{a}}(\sigma)(\log(Q_{\sf{a}}(\sigma)/\Lambda)-1)
\end{equation}
where $\alpha>0$ denotes the set of positive roots of $G$, the sum $\sum_{\sf{a}}$ is over the set of all weights of $\rho$ and $\Lambda$ is a UV energy cut-off scale\footnote{Note, in the equation \eqref{genericWeff}, $\sigma$ has units of energy, (in natural units).}. Then, the generic component of the discriminant loci $\Delta\subset \exp(\mathfrak{z}^{\vee}_{\mathbb{C}})$ in the $t$-space can be computed by the system of algebraic equations:
\begin{equation}\label{genericWeffeqs}
\exp\left(\frac{\partial\widetilde{W}_{\mathrm{eff}}(\sigma)}{\partial \sigma_{a}}\right)=1,\qquad \sigma\in\mathfrak{t}_{\mathbb{C}}/W_{G},
\end{equation}
upon some choice of a basis for the vector space $\mathfrak{t}_{\mathbb{C}}$. Due to the condition on our GLSM models to be nonanomalous, the eqs. \eqref{genericWeffeqs} depends only on $t$, not in $\sigma$, however, it is well known that, there exists several models where $\Delta$ gets contributions from regions were the large VEV's $\langle \sigma\rangle$ are not generic \cite{Morrison:1994fr}. These are known as mixed Coulomb-Higgs branches. So, more precisely if we denote by $\Delta_{\mathrm{gen}}$ the solutions to \eqref{genericWeffeqs}, we have
\begin{equation}
\Delta_{\mathrm{gen}}\subseteq \Delta.
\end{equation}
Only for $G$ abelian, there exist an algorithmic way to compute these mixed components (and in some cases, they may not exist). For general $G$, we only have some explicit families of examples \cite{Hori:2006dk,Knapp:2021vkm,Lin:2024fpz,Lin:2026icu}. We will be interested only on the phase crossing between $\zeta_{0}\ll-1$ and $\zeta_{0}\gg 1$, while keeping $\zeta_{\alpha}\gg 1$ for $\alpha=1,\ldots, r$. Therefore, for our computations, besides $\Delta_{\mathrm{gen}}$, the only relevant mixed Coulomb-Higgs branch are the ones that break the subgroup $U(1)\subset G$. In the case $G$ abelian, where $\bbG=\mathbb{P}^{m_{1}-1}\times\cdots\times \mathbb{P}^{m_{r}-1}$, we have that such a mixed branch can only exist if for each $I=m+1,\ldots,K$, there always exist at least one $\alpha=1,\ldots,r$ such that $n^{(\alpha)}_{I}$ does not vanish. Otherwise, if there exists $i_{*}\in\{1,\ldots,r\}$ such that $n^{(\alpha_{*})}_{I}=0$ for all $I=m+1,\ldots,K$, we cannot break just the subgroup $U(1)\subset G$. If we do, then $\zeta_{\alpha_{*}}$ can only take values in $\mathbb{R}_{\geq 0}$ giving us an invalid configuration \cite{Morrison:1994fr}. Therefore, we need to consider the mixed Coulomb-Higgs branch arising when we break at least $U(1)\times U(k_{\alpha_{*}})\subset G$ (with $k_{\alpha_{*}}=1$). In the latter case, we still need to check if $\zeta_{\alpha_{*}}\gg 1$ is compatible with the equations derived from the effective twisted potential in this sector.\\

A refined but heuristic analysis of quantum-corrected Kahler moduli space $\calM_K$ in FI-$\theta$ parameters around a discriminant, which is independent of the dimension of target space, can be found in the section 2.1 and appendix A in \cite{Lin:2026icu} for readers. Nevertheless, it is conjectured that the non-transversal intersection between quantum discriminant and large volume limits (for instance, $D_\alpha=\{z_\alpha:=\exp(- t_\alpha)=0\}\subset\calM_K$) forms a complement of torus links in the neighborhood of intersection points, such that the monodromy in this neighborhood is encoded in the fundamental group of the moduli space around each discriminant as the complement of links. As a consequence, any universal GLSM monodromy as a loop in the moduli space admits a decomposition from the braid group action from the loop of each component of discriminant, while the later components admit simple descriptions as spherical twists \cite{halpern2016autoequivalences,Cota:2019cjx,Lin:2024fpz}. We will elaborate on this point on the splitting model in the next section.

\subsection{GLSM brane and window shift monodromy}

\begin{figure}[h]
\centering
\includegraphics[width=.6\textwidth]{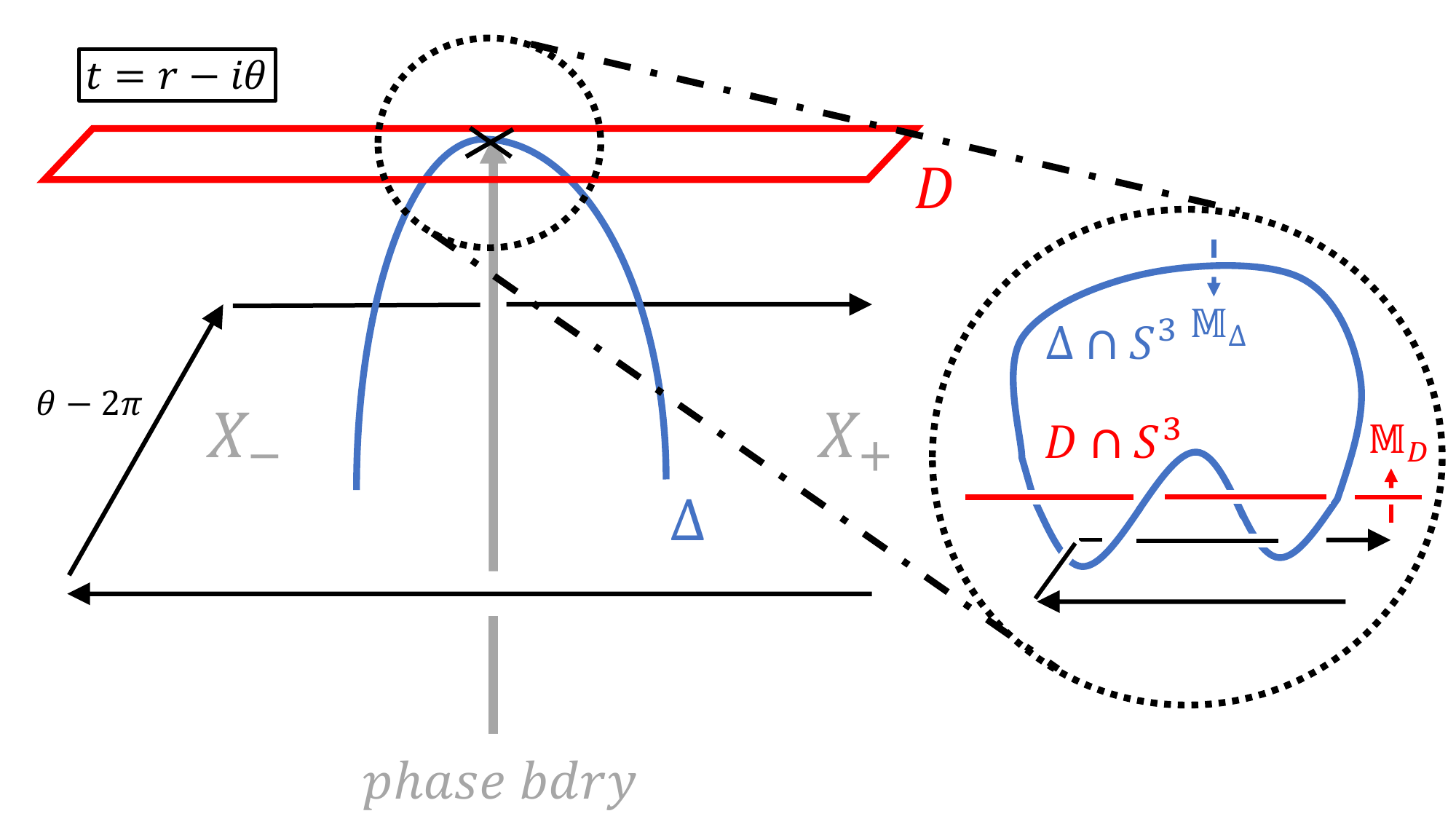}
\caption{Here the monodromy $M$ is illustrated by black arrows and $D:=\{z_{1}=0\}$. We write $\mathbb{M}_{\Delta}$ and $\mathbb{M}_{D}$ for the monodromies around the components  $S^{3}\cap D$ and $S^{3}\cap \Delta$, respectively.}\label{fig:windowshift2}
\end{figure}

B-type supersymmetric boundary conditions on GLSMs equipped with a superpotential $W$ are characterized by $G$-equivariant matrix factorizations of $W$, given by the triple $\calB:=(\mathbf{T},\rho_{M},R_{M})$\cite{Herbst:2008jq,Hori:2013ika} where $\mathbf{T}\in\mathrm{End}^{\mathrm{odd}}(M)$, is an odd endomorphism of the finite, $\mathbb{Z}_{2}$-graded, free $\mathrm{Sym} V^{\vee}$-module (it is customary to denote $M=M_{0}\oplus M_{1}$ for the even and odd factors of $M$). This triplet $\calB$ satisfies:
\begin{equation}
\begin{aligned}
    \mathbf{T}^2&=W\cdot\Id_{M},
    \\
    \rho_M^{-1}(g) \mathbf{T}(\rho(g)\cdot\phi)\rho_M(g)&= \mathbf{T}(\phi),\quad \text{for all \ }g\in G
    \\
    R_{M}(\lambda) \mathbf{T}(R(\lambda)\cdot\phi)R_{M}^{-1}(\lambda)&=\lambda \mathbf{T}(\phi),\quad\text{for all \ } \lambda\in U(1)_V.
\end{aligned}\label{eqn:MF}
\end{equation}
We will denote the category where B-branes belong, as $MF_{G}(W)$ (see \cite{ballard2019variation} for more details, in a more general case). When $X_{\zeta_+}$ is CY3, a set of such objects that correspond to holomorphic 0,2,4,6-cycles (D0,D2,D4,D6 branes) in $X_{\zeta_+}$ was constructed in the section 3.1 of \cite{Lin:2026icu} as
\begin{eqnarray}
    \{ \calB_{D0},\calB_{D2},\calB_{D4},\calB_{D6} \}=\{\mathcal{B}_{\mathrm{pt}},\mathcal{B}_{D_{\alpha}},\mathcal{B}_{\alpha},\widehat{\calE}_{-} \},\quad \alpha=0,1,\cdots,r.\label{eqn:Bbasis}
\end{eqnarray}
And we will discuss this basis for CY4 in next section. B-type boundary conditions on GLSMs need more data than just $\calB$. The boundary conditions for the vector multiplet must also be specified. The complete analysis of B-type boundary conditions in GLSMs, leads to the Grade Restriction Rule (GRR) \cite{Herbst:2008jq} and the definition of window categories (which are subcategories of $MF_{G}(W)$ \cite{ballard2019variation,halpern2015derived,segal2011equivalences}). We will not need to recall in detail the definition of the vector multiplet boundary condition \cite{Herbst:2008jq,Hori:2013ika} since we will be concerned with a very specific class of monodromies. The stringy K\"ahler moduli space $\mathcal{M}_{K}$ of a GLSM is the space spanned by the (exponentiated) FI-theta parameters $\exp t$. This space takes generically the form $(\mathbb{C}^{*})^{\mathrm{rk}\mathfrak{z}}\setminus \Delta$ and the GLSM provides a natural compactification of it \cite{Morrison:1994fr}. It is well known that $\mathcal{M}_{K}$ is subdivided into open chambers termed \emph{phases} \cite{Witten:1993yc}. For a nonanomalous GLSM, each point of $\mathcal{M}_{K}$ determines a SCFT (by RG flow to the IR fixed point), and B-type boundary conditions of a $\mathcal{N}=(2,2)$ SCFT form a triangulated category (more precisely a $A_{\infty}$ category, see for example \cite{Aspinwall:2009isa} for a review). If we denote this latter category $\mathcal{C}_{c}$ for $t$ in a given chamber $c$ of $\mathcal{M}_{K}$, we always have a projection functor:
\begin{equation}\label{projfunctor}
\pi_{c}:MF_{G}(W)\rightarrow \mathcal{C}_{c}.
\end{equation}
When a chamber $c$ corresponds to a geometric phase, where the IR fixed point corresponds to the fixed point of a sigma model with target space $X$, we have
\begin{equation}
\mathcal{C}_{c}\cong D(X):=D^{b}\mathrm{Coh}(X),\qquad c \text{ \ is geometric}.
\end{equation}
The image of projection on \eqref{eqn:Bbasis} are coherent sheaves as (twisted) structure sheaves on point class $\calO_P$, twisted curve classes\footnote{We emphasis that the curve class is twisted such that $\calO_{\widetilde{C_\alpha}}\cong \calO_{\bbP^1}(-1)$.} $\calO_{\widetilde{C_\alpha}}$, divisor classes $\calO_{D_\alpha}$ and structure sheaf $\calO_{X_{\zeta_+}}$. Given a point in the covering $\widetilde{\mathcal{M}}_{K}\rightarrow\mathcal{M}_{K}$, defined simply by the logarithmic map i.e. the coordinates $t\in\widetilde{\mathcal{M}}_{K}$ are just defined by 'unwrapping' the $\theta$ coordinate, allowing it to take values in $\mathbb{R}$. A point in $\widetilde{\mathcal{M}}_{K}$ determines a subcategory of $\mathbb{W}\subset MF_{G}(W)$ called a window category. In the following, we will describe it for the cases of interest. Consider the PAX model and the phases:
\begin{equation}\label{pmphases}
\zeta_{\pm}:=(\pm\zeta_{0}\gg 1,\zeta_{1}\gg 1,\ldots,\zeta_{r}\gg 1)
\end{equation}
then, the window subcategory defined by a straight line running between phases $\zeta_{\pm}$ is defined by restricting the weights of $\rho_{M}$ to the band\footnote{It is called a 'band'\cite{Herbst:2008jq} because the rest of the weights are unrestricted.}:
\begin{equation}\label{abGRR}
-\frac{m}{2}<q^{0}+\frac{\theta^{0}}{2\pi}<\frac{m}{2}
\end{equation}
where $q^{0}$ denotes the weight of the $U(1)\subset G$ subgroup. Then the window subcategories, became, explicitly:
\begin{equation}\label{windowabdef}
\mathbb{W}(l):=\{\mathcal{B}\in MF_{G}(W):\text{weights of \ } \rho_{M}\text{ \ satisfy \eqref{abGRR} }\},\qquad
l:=\left\lfloor{\frac{\theta^{0}}{2\pi}}\right\rfloor.
\end{equation}
When $\mathrm{dim}( \calM_K)=2$, denote the coordinate as $z_0$ and $z_1$, the monodromy giving rise to the window categories $\mathbb{W}(l)$ corresponds to loops running parallel to $\{z_{1}=0\}\cap S^{3}$, surrounding intersection point. Therefore we expect that the autoequivalence $M\in \mathrm{Aut}(D(X_{\zeta_+}))$ corresponding to this monodromy can be decomposed in a composition of other elements of $\mathrm{Aut}(D(X_{\zeta_+}))$ consistent with the relations of $\pi_1(S^3\setminus\mathfrak{L})$ where $\mathfrak{L}=S^{3}\cap (\{z_{1}=0\}\cup \Delta)$. This is sketched in figure~\ref{fig:windowshift2}.

Now we have all set to compute the monodromy functor associated to a loop, surrounding the phase boundary corresponding to the geometric phases $X_{\zeta_{+}}$ and $X_{\zeta_{-}}$ in $\mathcal{M}_{K}$ with base point near $\zeta_{+}$. To be more precise, we take a loop that keeps $z_{\alpha}=\exp(-t_{\alpha})=\varepsilon_{\alpha}=\mathrm{const.}$, $\alpha=1,\ldots,r$ fixed at $|\varepsilon_{i}|\ll 1$ for all $i$ and we only vary $z_{0}$ on a loop surrounding the points
\begin{eqnarray}
\Delta\cap\bigcap_{i=1}^{r}\{ z_{i}=\varepsilon_{i}\}\subset \mathcal{M}_{K} 
\end{eqnarray}
once. This loop is sketched, for the case 
$\mathrm{dim}\mathcal{M}_{K}=2$ in figure
\ref{fig:windowshift2}. We can then fix $\theta_{0}$ so that the corresponding window categories \eqref{windowabdef} for the loop corresponds to the following sequence of equivalences:
\begin{eqnarray}
F_{\mathcal{E}_{-}}&:&D(X_{\zeta_{+}})\cong \mathbb{W}(0)\rightarrow\mathbb{W}(-1),\nonumber\\
F_{\mathcal{E}_{+}}&:&\mathbb{W}(-1)\rightarrow\mathbb{W}(0)\cong D(X_{\zeta_{+}})
\end{eqnarray}
where $F_{\mathcal{E}_{\pm}}$ corresponds to taking cones with (twists of) $\mathcal{E}_{\pm}\in MF_{G}(W)$, the exact objects in the kernel of projection functor ~\eqref{projfunctor}. This will result on the monodromy functor:
\begin{eqnarray}
M:=F_{\mathcal{E}_{+}}\circ F_{\mathcal{E}_{-}}\in\mathrm{Aut}(\mathbb{W}(0))\cong \mathrm{Aut}(D(X_{\zeta_{+}})).
\end{eqnarray}

We face the problem that not all the objects in \eqref{eqn:Bbasis} are in $\mathbb{W}(0)$ i.e. we need to grade restrict them to $\mathbb{W}(0)$. This is possible by applying the projection functor $\pi_c$ by taking multiple cones with $\mathcal{E}_{+}$ and their twists. This computation on the basis \eqref{eqn:Bbasis} is comprehensively analyzed in \cite{Lin:2026icu}, and some CY4 results can also be read directly from there. Nevertheless, we will continue the  monodromy analysis for CY4 in section \ref{sec:monodromy}.

\section{A-period lattice for CY4}\label{sec:period}

The central charge of B-branes for $\mathcal{N}=(2,2)$ theories is defined in \cite{Cecotti:1991me,hori2000d} by the partition function the A-twisted theory on a disk attached to an infinitely flat cylinder and boundary conditions corresponding to a B-brane. This coupling of an A-twisted theory in the bulk and B-brane boundary conditions is a very natural object to study, for instance in $\mathcal{N}=(2,2)$ SCFTs \cite{Ooguri:1996ck}. Using supersymmetric localization techniques 
\cite{Hori:2013ika,Honda:2013uca,Sugishita:2013jca} the 
central charge for a B-brane $(\mathcal{B},L_{t})$, $\mathcal{B}\in MF_{G}(W)$ was computed in the context of GLSMs and found to be given by a Mellin-Barnes type integral:
\begin{equation}
Z_\mathcal{B}(t):=\int_{L_{t}\subset 
\mathfrak{t}_{\bbC}} \mathrm{d}^{l_{G}}\sigma 
\prod_{\alpha>0}\alpha(\sigma)\sinh(\pi\alpha(\sigma))\prod_{j=1}^{\mathrm{rk}V}\Gamma\left(iQ_{j}(\sigma)+\frac{R_{j}}{2}\right)e^{it(\sigma)}f_{\mathcal{B}}(\sigma).
\end{equation}
where $Q_j\in \mathfrak{t}^{\vee}_{\bbC}$ denotes the weights of the representation \eqref{eqn:chiralfields}, while $R_j$ denotes the weights of the $U(1)_R$ symmetry action. The symbol $\prod_{\alpha>0}$ denotes the product over the positive roots of $G$ and $l_{G}:=\mathrm{dim}(\mathfrak{t})$. Denote $\sigma^{(i)}_\alpha$ the coordinates of $(\mathfrak{t}_{U(k_\alpha)})_{\bbC}\cong \mathbb{C}^{k_{\alpha}}$, $i=1,\cdots,k_\alpha$, and define
\begin{equation}
\sigma_\alpha:=\sigma_\alpha^{(1)}+\cdots+\sigma_\alpha^{(k_\alpha)},\qquad,\alpha=0,\ldots, r.
\end{equation}
In particular, $k_0=1$ and $m_0=m$. For the GLSM in \eqref{eqn:chiralfields}, the hemisphere partition function, with respect to the R-charge integrality \eqref{eqn:RInt}, has an explicit expression as
\begin{eqnarray}
  Z_{\calB}(t) &=& \int_{L_t} \prod_{\alpha=0}^r \mathrm{d}^{k_{\alpha}}\sigma_\alpha\prod_{i_\alpha<j_\alpha}(\sigma_\alpha^{(i_\alpha)}-\sigma_\alpha^{(j_\alpha)})\sinh\pi(\sigma_\alpha^{(i_\alpha)}-\sigma_\alpha^{(j_\alpha)}) e^{it_\alpha \sigma_\alpha}\prod_{i_\alpha=1}^{k_\alpha}\Gamma(i\sigma^{(i_\alpha)}_\alpha)^{m_\alpha}
  \nonumber\\
  &&\times \prod_{I=1}^{m}\Gamma\left(-i \vec n_I(\sigma)-i\sigma_0+1\right)\times \prod_{J=m+1}^{K}\Gamma(-i \vec n_J(\sigma)+1) f_{\calB}(\sigma).
  \nonumber\\ \label{eqn:ZBt}
\end{eqnarray}
Where we write:
\begin{eqnarray}
\vec{n}_I(\sigma):=\sum_{\alpha=1}^{r}n_{I}^{(\alpha)}\sigma_{\alpha},\qquad I=1,\ldots,K.
\end{eqnarray}
Finally, the contribution of the object $\calB$ is contained in the brane factor $f_{\calB}$ and is given by:
\begin{equation}
f_{\mathcal{B}}(\sigma):=\mathrm{tr}_{M}\left(R_{M}(e^{i
\pi})\rho_{M}(e^{2\pi\sigma
} )\right)
\end{equation}

Since we are interested in the monodromies with base point near $\zeta_{+}$, $Z_{\mathcal{B}}(t)$ can be written as the infinite sum of residues:
\begin{eqnarray}\label{resLplus}
  Z_{\calB}(t)|_{\zeta_{+}} = \sum_{il_{\alpha}\in i(\mathbb{Z}_{\geq 0})^{k_{\alpha}}}\oint_{s=0}\prod_{\alpha=0}^r \mathrm{d}^{k_{\alpha}}s_{\alpha} h_{\mathcal{B}}(s^{(i_{\alpha})}_{\alpha}+il^{(i_{\alpha})}_{\alpha}),
\end{eqnarray}
where $h_{\calB}(\sigma_{\alpha}^{(i_{\alpha})})$ denotes the integrand of \eqref{eqn:ZBt}. Since the poles of the gamma functions are simple poles, the integral \eqref{resLplus} can be straightforwardly identified with an integral over $\mathbb{P}^{m-1}\times (\mathbb{P}^{m_{1}-1})^{k_{1}}\times\cdots\times (\mathbb{P}^{m_{r}-1})^{k_{r}}$. Moreover, as it is done in \cite{Lin:2024fpz}, using the results of \cite{Martin:1999ng}, we can write \eqref{eqn:ZBt} as an integral over $\mathbb{P}^{m-1}\times \bbG$, indeed we can state the following proposition:
\begin{prop}\cite{bertram1996severi}\label{propresgrass}
    Denote $H^{(1)},\cdots,H^{(k)}$ the Chern roots of the dual tautological bundle $\mathcal{S}^{\vee}\rightarrow G(k,m)$. Then for any regular totally-symmetric function $h(x)$ of the variables $x^{(1)},\cdots,x^{(k)}$, we have the identity
    \begin{equation}
        \frac{(-1)^{\tiny\begin{pmatrix}
            k\\2
        \end{pmatrix}}}{k!}\oint_0 \frac{\mathrm{d}^k x}{(2\pi i)^k}  \ \frac{\prod_{i<j}(x^{(i)}-x^{(j)})^2}{(x^{(1)}\cdots x^{(k)})^m}  h(x)=\int_{G(k,m)} h(H^{(i)}).
    \end{equation}
\end{prop}
Thus, the residues \eqref{resLplus} become an integral over $\mathbb{P}^{m-1}\times \bbG$ upon the identification
\begin{equation}\label{sigmaident}
   s_\alpha^{(i_\alpha)} \rightarrow \frac{H_\alpha^{(i_\alpha)}}{2\pi}, \qquad c(\mathcal{S}^{\vee}_{\alpha})=\prod_{i_{\alpha}=1}^{k_{\alpha}}(1+H_{\alpha}^{(i_{\alpha})}).
\end{equation}
The integration over $\mathbb{P}^{n}\times \bbG$ can be further reduced to an integration over $X_{\zeta_{+}}$ by use of the adjunction formula. It is then expected that \eqref{resLplus} reduces to the geometric central charge of B-branes or A-periods \cite{Cheung:1997az,Green:1996dd,minasian1997k}, in the $\zeta_{+}$ phase. More precisely we expect that the geometric A-period of a sheaf $\mathcal{E}\in D(X_{\zeta_{+}})$, $Z^{\mathrm{geom}}_{\mathcal{E}}$, coincides with \eqref{resLplus} as
\begin{equation}\label{conjecturegeom}
Z_{\mathcal{B}}|_{\zeta_{+}}=Z^{\mathrm{geom}}_{\pi_{\zeta_{+}}(\mathcal{B})}
\end{equation}
in the following we will simply denote it by $Z_{\mathcal{E}}$, since we will not refer to A-periods in any other phase. Moreover, we will work on the particular basis of A-periods that will be discussed momentarily.\\

As a conclusion, the A-period $Z_{\mathcal{E}}$ of $X$, in general, takes the form: \cite{Gerhardus:2016iot,Lin:2026icu}
\begin{equation}
\begin{gathered}
Z_{\mathcal{E}}(\kappa)=\int_X 
e^{J}\hat\Gamma_X\mathrm{ch}(\calE)+\text{instantons}=:Z^{0}_{\calE}(\kappa)+\text{instantons } ,
\end{gathered}\label{LVZ}
\end{equation}
where $J:=B+i\frac{\omega}{2\pi}\in H^2(X,\mathbb{C})$, $B\in 
H^{2}(X,\mathbb{R}/\mathbb{Z})$ is the $B$-field and $\omega\in 
\mathcal{K}_{X}\subset H^{2}(X,\mathbb{R})$. $\mathcal{K}_{X}$ denotes the 
K\"ahler cone of $X$ and the instantons are weighted by
\begin{equation}\label{instantonexp}
\exp\left(2\pi i\int_{\beta}J\right),\quad \beta\in H_{2}(X,\mathbb{Z})
\text{ an effective curve class.}
\end{equation}
Fix a basis $\{J_{\alpha}\}$ of $H^2(X,\bbC)$, then $J=\kappa^\alpha J_\alpha$, and $\kappa^{\alpha}$ corresponds to the so-called flat coordinates. The Gamma class\footnote{ For $X$ CY, the $\hat{A}$-genus 
equals the Todd class $\mathrm{Td}_{X}$. Because of (\ref{gammaroot}), we can 
regard $\hat\Gamma_X$ as a root of $\hat{A}_{X}$. Moreover, this root is not unique and $\hat\Gamma_X$ is just a particular choice \cite{Halverson:2013qca}, however this choice is different from the one corresponding to the 
Ramond-Ramond (RR) charge computed in \cite{Cheung:1997az,Green:1996dd,minasian1997k}. The choice $\hat\Gamma_X$ encodes the perturbative corrections to the central charge. The Gamma class appears implicitly in the works 
\cite{libgober1999chern,Hosono:2000eb} and then it was further defined in a mathematical context in \cite{iritani2009integral,katzarkov2008hodge}.}
$\hat\Gamma_X$ is a multiplicative characteristic class, given by 
\begin{equation}
\hat\Gamma_X:=\prod_{j}\Gamma\left(1-\frac{\lambda_{j}}{2\pi i}\right)
\end{equation}
where $\lambda_{j}$ are the Chern roots of the holomorphic tangent bundle $TX$ 
of 
$X$. It satisfies the important property
\begin{equation}\label{gammaroot}
\hat\Gamma_X\hat{\Gamma}^{*}_{X}=\hat{A}_{X},\qquad \hat\Gamma^{*}_X:=\prod_{j}\Gamma\left(1+\frac{\lambda_{j}}{2\pi i}\right)
\end{equation}

\subsection{Homology cycles and their classical A-periods}

From the discussion above, we can only work on the classical period $Z_\calB$ in the integral form \eqref{LVZ}, for $X\subset\bbP^{m-1}\times G(k_1,m_1)\times\cdots\times G(k_r,m_r)$, presumably $k_\alpha=1,2$, as a complete intersection \eqref{eqn: CICYI}. We shall establish our notations here for the following computation. We denote $\calS_\alpha^\vee$ the tautological dual bundle of each $G(k_\alpha,m_\alpha)$ and its Chern roots as $H_\alpha^{(1)},\cdots, H_\alpha^{(k_\alpha)}$, thus by splitting principle
\begin{eqnarray}
    c(\calS_\alpha^\vee)=\prod_{i_\alpha=1}^{k_\alpha}(1+H_\alpha^{(i_\alpha)})
\end{eqnarray}
We define 
\begin{eqnarray}
    H_\alpha:=c_1(\calS_\alpha^\vee)=H_\alpha^{(1)}+\cdots+ H_\alpha^{(k_\alpha)}
\end{eqnarray}
and
\begin{eqnarray}
    \widetilde{H}_\alpha:=c_{k_\alpha}(\calS_\alpha^\vee)=H_\alpha^{(1)}\cdots H_\alpha^{(k_\alpha)}.
\end{eqnarray}
They are normalized as
\begin{equation}
    \int_{\bbG}\prod_{\alpha=1}^r \widetilde{H}_\alpha^{m_\alpha-k_\alpha}=1.\label{eq_norm}
\end{equation}

The Gamma class and Todd class for CY4 $X$ is explicitly given by
\begin{equation}
    \Gamma(X)=1+\frac{1}{24}c_2(X)+\frac{\zeta(3)}{(2\pi i)^3}c_3(X)+\frac{1}{5760}\left(7c_2(X)^2-4c_4(X)\right)
\end{equation}
and 
\begin{equation}
    \Td(X)=\Gamma(X)\overline{\Gamma}(X)^\vee=1+\frac{1}{12}c_2(X)+\frac{1}{720}(3c_2(X)^2-c_4(X)).
\end{equation}
Note that by the Hirzebruch-Riemann-Roch formula, the Euler number for spherical object $\calO_X$ is
\begin{equation}
    \frac{1}{720}\int_X 3c_2^2 -c_4=\chi(\calO_X)\equiv2.
\end{equation}
Thus under the integration over $X$, the top degree part in $\Gamma(X)$ is always as
\begin{eqnarray}
     \int_X\frac{1}{5760}\left(7c_2(X)^2-4c_4(X)\right)\equiv\frac{7}{12}-\frac{1}{3456}\int_X c_4(X).
\end{eqnarray}
The classical part of the hemisphere partition function (classical A-period) is now explicitly given by
\begin{eqnarray}
    Z_{\calE}(\kappa)&=&\int_Xe^{J} \Gamma(X)\Ch(\calE)
    \nonumber\\
    &=&\int_X\Ch_0(\calE)\left( \frac{J^4}{4!}+ \frac{c_2J^2}{48}+\frac{\zeta(3)}{(2\pi i)^3}c_3\cdot J+\frac{7c_2^2-4c_4}{5760}\right)
    \nonumber\\
    &&+\Ch_1(\calE)\left(\frac{J^3}{3!}+\frac{c_2\cdot J}{24}+\frac{\zeta(3)}{(2\pi i)^3}c_3\right)
    \nonumber\\
    &&+\Ch_2(\calE)\left( \frac{J^2}{2}+\frac{c_2}{24} \right)+\Ch_3(\calE)\cdot J+\Ch_4(\calE)
\label{centrexp}
\end{eqnarray}

Now we review the charge basis for CY4 $X$ as complete intersection in grassmannian in \cite{Gerhardus:2016iot}. A Doran-Morgan basis consists of objects that are dual to holomorphic cycles in the lattice of integral Hodge classes, $H^{\mathrm{ev}}(X,\bbZ)\cap H^{*,*}(X)$:
\begin{equation}
    \langle D8,\ D6_\alpha,\ \{D4_{\Gamma}\},\ D2_\alpha,\ D0 \rangle=\langle\; \calO_X,\ \calO_{D_\alpha},\ \{ \calO_{S_{\alpha\beta}},\ \calO_{S_\alpha}\},\ {\calO_{\widetilde{C_\alpha}}},\ \calO_P\;\rangle
\end{equation}
Notice that there are the following subtleties in this charge basis. First of all, if $G(k_\alpha,m_\alpha)\cong\bbP^1$, then $\calO_{S_{\alpha\alpha}}$ will be empty since $[S_{\alpha\alpha}]\cong\bbP^2$ does not exist in $G(k_\alpha,m_\alpha)$. Also, $\calO_{S_\alpha}$ does not exist if $k_\alpha=1$. Moreover, it is not clear how to geometrically construct $\calO_{S_\alpha}$ if $k_\alpha=\rank\calS_\alpha\geq3$, although a direct analog can be written down at the level of $c_2(\calS_\alpha)$. We will avoid the case-by-case discussion for simplicity.\\

The construction of corresponding B-brane objects is parallel to the section 3.1 of \cite{Lin:2026icu}. The $D0$ brane locates on a point in $X$ corresponds to a skyscraper sheaf $\calO_P$ for $P$ a point on $X$. One may in turn view it as arising from $\bbG$ via complete intersection of zeros of sections of the following bundle, in accord with \eqref{eq_norm},
\begin{eqnarray}
    (\calS_1^\vee)^{\oplus(m_1-k_1)}\boxtimes\cdots\boxtimes(\calS_r^\vee)^{\oplus(m_r-k_r)}.
\end{eqnarray}
where $\boxtimes$ denotes the external tensor product, with bundles pulled back from $G(k_\alpha,m_\alpha)$ to $A$. Restricting $H_\alpha$ to $X$, we choose a basis with $r$ generators for $H^2(X,\bbC)$ as
\begin{eqnarray}
    J=\sum_{\alpha=1}^r \kappa^\alpha J_\alpha=\sum_{\alpha=1}^r \kappa^\alpha\ i^*H_\alpha.
\end{eqnarray}
We take the divisor classes on $X$ that represent these cohomology classes as $D_\alpha$. A divisor class that a $D6$ brane wraps on fits into the exact sequence
\begin{eqnarray}
    0\rightarrow\calO_X(-J_\alpha)\rightarrow\calO_X\rightarrow\calO_{D_\alpha}\rightarrow 0.
\end{eqnarray}
By Poincar\'e duality, there are as many as $r$ curve classes in $H^6(X,\bbC)$. Similarly, they come from restricting to $X$ the lines $C_\alpha\subset \bbG$ given by the following complete intersection, which are themselves $\bbP^1$ \footnote{Note that $k(m-k-1)$ sections in the dual tautological bundle fix $G(k,m)$ to $G(k,k+1)\cong \bbP^k$, then $(k-1)$ sections in the dual tautological line bundle fix $\bbP^k$ to $\bbP^1$. By counting the degree of freedom, we assume that the choice of sections generally exists.}
\begin{eqnarray}
\big((\calS_\alpha^\vee)^{\oplus(m_\alpha-k_\alpha-1)}\oplus (\det\calS_\alpha^{\vee})^{\oplus(k_\alpha-1)}  \big)\boxtimes\bigboxtimes_{\beta\neq\alpha}(\calS_\beta^\vee)^{\oplus(m_\beta-k_\beta)}.
\end{eqnarray}
The $D2$ branes in our basis wrap on twisted curves $\widetilde{C_\alpha}$ with $\calO_{\widetilde{C_\alpha}}\cong\calO_{\bbP^1}(-1)$. Finally, the four-cycles on $X$ that $D4$ branes wrap on come from the pull back of $H^4(A,\bbC)$ that contains $h^{2,2}(X)$ generators, essentially spanned by $c_2(\calS_\alpha^\vee)$ and $c_2(\calS^\vee_\alpha\boxtimes\calS^\vee_{\beta})=c_1(\calS^\vee_\alpha)c_1(\calS^\vee_{\beta})$. They are defined by the following exact sequences:
\begin{eqnarray}
    0\rightarrow \calO_X(-J_\alpha-J_\beta)\rightarrow\calO_X(-J_\alpha)\oplus\calO_X(-J_\beta)\rightarrow\calO_X\rightarrow\calO_{S_{\alpha\beta}}\rightarrow 0
\end{eqnarray}
and
\begin{eqnarray}
     0\rightarrow \calO_X(-J_\alpha)\rightarrow\calS_\alpha|_X\rightarrow\calO_X\rightarrow\calO_{S_{\alpha}}\rightarrow 0.
\end{eqnarray}

The Chern characters of the above cycles are (for $J_\mu^{(j)}:=i^*H_\mu^{(j)}$ and $\widetilde J_\mu:=i^*(H_\alpha^{(1)}H_\alpha^{(2)})$)
\begin{eqnarray}
    \Ch\calO_X&=&1
    \\
    \Ch\calO_{D_\mu}&=&1-e^{-J_\mu}
    \nonumber\\
    &=&J_\mu-\frac{1}{2}J_\mu^2+\frac{1}{3!}J_\mu^3-\frac{1}{4!}J_\mu^4
    \\
    \Ch \calO_{S_{\mu\nu}}&=&(1-e^{-J_\mu})(1-e^{-J_\nu})
    \nonumber\\
    &=&J_\mu J_\nu\left(1-\frac{J_\mu+J_\nu}{2}+\frac{J_\mu^2}{6}+\frac{J_\mu J_\nu}{4}+\frac{J_\nu^2}{6}\right)
    \\
    \Ch\calO_{S_\mu}&=&(1-e^{-J_\mu^{(1)}})(1-e^{-J_\mu^{(2)}})
    \nonumber\\
    &=&\widetilde{J}_\mu\left(1-\frac{J_\mu}{2}+\frac{J_\mu^2}{6}-\frac{\widetilde{J}_\mu}{12}\right)
    \\
    \Ch\calO_{\widetilde{C_{\mu}}}&=&\frac{e^{-H_\mu}}{f_{\calO_X}(H)} (1-e^{-H_\mu})^{k_\mu-1}\times\prod_{\alpha=1}^r\prod_{i_\alpha=1}^{k_\alpha}(1-e^{-H^{(i_\alpha)}_\alpha})^{m_\alpha-k_\alpha-\delta_{\mu\alpha}}\label{eq_chC}
    \\
    \Ch\calO_P&=&\frac{1}{f_{\calO_X}(H)}\times \prod_{\alpha=1}^r\prod_{i_\alpha=1}^{k_\alpha}(1-e^{-H^{(i_\alpha)}_\alpha})^{m_\alpha-k_\alpha}\label{eq_chpt}
\end{eqnarray}
where there is a formal inverse of $f_{\calO_X}(H)$, which is
\begin{equation}
    f_{\calO_X}(H)=\prod_{I=1}^{K}\left(1-e^{- r_I}\right).
\end{equation}

It is straightforward to expand $Z_{\calE}$ for $\calE=\calO_X$, $\calO_{D}$ and $\calO_{S}$ as
\begin{eqnarray}
    Z_{X}(\kappa)&=&\frac{1}{4!}c_{\alpha\beta\rho\sigma}\cdot(\kappa^4)+\frac{1}{2}c_{\alpha\beta}\kappa^\alpha\kappa^\beta+c_\alpha \kappa_\alpha+\frac{7}{12}-\frac{1}{3456}\chi(X)
\\
     Z_{{D_\mu}}(\kappa)&=& \frac{1}{3!}c_{\mu\alpha\beta\gamma}\cdot (\kappa^3)-\frac{1}{4}c_{\mu\mu\alpha\beta}\kappa^\alpha\kappa^\beta+\left(\frac{1}{6}c_{\mu\mu\mu\alpha}+c_{\mu\alpha}\right)\kappa^\alpha
    \nonumber\\
    &&-\frac{1}{4!}c_{\mu\mu\mu\mu}-\frac{1}{2}c_{\mu\mu}+c_\mu
\\
    Z_{S_{\mu\nu}}(\kappa)&=&\frac{1}{2}c_{\mu\nu\alpha\beta}\kappa^{\alpha}\kappa^\beta-\frac{1}{2}\left(c_{\mu\mu\nu\alpha}+c_{\mu\nu\nu\alpha}\right)\kappa^\alpha
    \nonumber\\
    &&+\frac{1}{6}c_{\mu\mu\mu\nu}+\frac{1}{4}c_{\mu\mu\nu\nu}+\frac{1}{6}c_{\mu\nu\nu\nu}+c_{\mu\nu},
\\
    Z_{S_\mu}(\kappa)&=& \frac{1}{2}\widetilde{c_{\mu\alpha\beta}}\kappa^\alpha\kappa^\beta - \frac{1}{2} \widetilde{c_{\mu\mu\alpha}} \kappa^\alpha +\frac{1}{6}\widetilde{c_{\mu\mu\mu}}-\widetilde{c_{\mu\mu}}+\widetilde{c_\mu}.
\end{eqnarray}
The coefficients in the above are given by topological intersection numbers as
\begin{eqnarray}
    c_{\alpha\beta\mu\nu}&=&\int_X J_\alpha J_\beta J_\mu J_\nu,\quad c_{\alpha\beta}=\frac{1}{24}\int_X c_2(X)J_\alpha J_\beta,\quad c_\alpha={\zeta(3)\over (2\pi i)^3}\int_X c_3(X)J_\alpha,
    \nonumber\\
    \widetilde{c_{\mu\alpha\beta}}&=& \int_X \widetilde{J}_\mu J_\alpha J_\beta,\quad \widetilde{c_{\mu\mu}}=\frac{1}{12}\int_X\widetilde{J}_\mu^2,\quad \widetilde{c_\mu}=\frac{1}{24}\int_X  c_2(X)\widetilde{J}_\mu.
\end{eqnarray}
The classical A-periods for $\calO_{\widetilde{C_\alpha}}$ and $\calO_P$ are given by the following as in CY3 cases \cite{Lin:2026icu} (We will show a similar derivation on dual curve classes momentarily)
\begin{align}
    Z_P(\kappa)=1, 
    \\
    Z_{\widetilde {C_\mu}}(\kappa)=\kappa_\mu.
\end{align}

Notice that $\chi$ is symmetric for CY4 $X$. The pairing matrix can be computed using either the annulus partition function for a GLSM geometric phase or the Hirzebruch-Riemann-Roch formula. The symmetric pairing matrix on our basis has following entries:
\begin{eqnarray}
    \chi(\calO_X,\calO_X)&=& 2
    \\
    \chi(\calO_X,\calO_{D_\alpha})&=& -\frac{1}{24} c_{\alpha\alpha\alpha\alpha}- c_{\alpha\alpha}
    \\
    \chi(\calO_X,\calO_{S_{\alpha\beta}})&=&\frac{1}{6}c_{\alpha\beta\beta\beta}+\frac{1}{4}c_{\alpha\alpha\beta\beta}+\frac{1}{6}c_{\alpha\alpha\alpha\beta}+2c_{\alpha\beta}
    \\
    \chi(\calO_X,\calO_{S_\alpha})&=&\frac{1}{6}\widetilde{c_{\alpha\alpha\alpha}}-\widetilde{c_{\alpha\alpha}}+2\widetilde{c_\alpha}
    \\
    \chi(\calO_X,\calO_P)&=& 1
\end{eqnarray}
and
\begin{eqnarray}
     \chi(\calO_{D_\alpha},\calO_{D_\beta})&=&-\chi(\calO_X,\calO_{S_{\alpha\beta}})+\frac{1}{2}c_{\alpha\alpha\beta\beta}
     \\
     \chi(\calO_{D_\alpha},\calO_{S_{\beta\gamma}})&=& \frac{1}{2}(c_{\alpha\beta\gamma\gamma}+c_{\alpha\beta\beta\gamma}-c_{\alpha\alpha\beta\gamma})
     \\
    \chi(\calO_{D_\alpha},\calO_{S_\beta})&=&\frac{1}{2}(\widetilde{c_{\beta\beta\alpha}}-\widetilde{c_{\beta\alpha\alpha}})
     \\
     \chi(\calO_{D_\alpha},\calO_{\widetilde{C_\beta}})&=&-\delta_{\alpha\beta}
\end{eqnarray}
and finally
\begin{eqnarray}
    \chi(\calO_{S_{\alpha\beta}},\calO_{S_{\rho\sigma}})&=& c_{\alpha\beta\rho\sigma}
    \\
    \chi(\calO_{S_{\alpha\beta}},\calO_{S_\rho})&=&
\end{eqnarray}
while all other pairings vanish.

\subsection{Dual surface classes}

Now we introduce $\calO_{\widetilde{S_{\alpha\beta}}}$, the dual class of $\calO_{S_{\alpha\beta}}$ in $H^4(X,\bbC)$, satisfying
\begin{eqnarray}
    \chi(\calO_{S_{\alpha\beta}},\calO_{\widetilde{S_{\rho\sigma}}})=\delta_{\alpha\rho}\delta_{\beta\sigma}
\end{eqnarray}
with algebraic cycles represented by restricting to $X$ a surface class equivalent to $\bbP^2$ or $\bbP^1\times\bbP^1$ in $\bbG$, that will be directly related to GLSM monodromy. Similarly to the construction for curve classes, the dual surface class $\widetilde{S_{\alpha\beta}}$ is given by a surface in $\bbG$ that is homologous to a Hirzebruch surface $\bbP^1\times\bbP^1$ (resp. rational surface $\bbP^2$) when $\mu\neq\nu$ (resp. $\mu=\nu$) from the complete intersection of the following bundle on $\bbG$:
\begin{eqnarray}
    {\bigboxtimes}_{\rho=1}^r\bigg\{(\calS_\rho^\vee)^{\oplus(m_\rho-k_\rho-s_{\alpha\beta}(\rho))}\oplus (\det\calS_\rho^{\vee})^{\oplus(k_\rho-\delta_{\rho\alpha}-\delta_{\rho\beta})\varepsilon_{\alpha\beta}(\rho)}  \bigg\}.
\end{eqnarray}
in where function $\varepsilon_{\alpha\beta}(\rho)$ is
\begin{eqnarray}
    \varepsilon_{\alpha\beta}(\rho)=\delta_{\rho\alpha}+\delta_{\rho\beta}-\delta_{\rho\alpha}\delta_{\rho\beta}=\left\{\begin{array}{cc}
        1 &  \rho=\alpha \text{ or }\beta,
         \\
        0 &  \text{otherwise}.
    \end{array}\right.
\end{eqnarray}
Moreover, we twist the structure sheaves by appropriate Chan-Paton bundles
\begin{eqnarray}
    \calO_{\widetilde{S_{\mu\nu}}}:=\calO_{\bbP^1\times\bbP^1}(-1,-1),\qquad \calO_{\widetilde{S_{\mu\mu}}}:= \calO_{\bbP^2}(-2).
\end{eqnarray}
The Chern character is hence given by
\begin{eqnarray}
     \Ch\calO_{\widetilde{S_{\mu\nu}}}&=&\frac{e^{-H_\mu-H_\nu}}{f_{\calO_X}(H)}\times \prod_{\alpha=1}^r(1-e^{-H_\alpha})^{(k_\alpha-\delta_{\alpha\mu}-\delta_{\beta\nu})\varepsilon_{\mu\nu}(\alpha)}
     \nonumber\\
     &&\times\prod_{i_\alpha=1}^{k_\alpha}(1-e^{-H^{(i_\alpha)}_\alpha})^{(m_\alpha-k_\alpha-\varepsilon_{\mu\nu}(\alpha))}.
\end{eqnarray}
For later computation, the numerator can be expanded as follows: for $\mu\neq\nu$, 
\begin{eqnarray}
   f_{\calO_X} \Ch\calO_{\widetilde{S_{\mu\nu}}}&=&\prod_{\alpha=\mu,\nu}\bigg(e^{-H_\alpha}(1-e^{-H_\alpha})^{k_\alpha-1} \times\prod_{i_\alpha=1}^{k_\alpha}(1-e^{-H^{(i_\alpha)}_\alpha})^{m_\alpha-k_\alpha-1}\bigg)\times\prod_{\beta\neq\mu,\nu} H_\beta^{(1)}\cdots H_\beta^{(k_\beta)}
    \nonumber\\
    &=& \prod_{\alpha=\mu,\nu} \bigg( \widetilde{H}_\alpha^{m_\alpha-k_\alpha-1}\left( -\frac{m_\alpha}{2}H_\alpha^{k_\alpha}+H_\alpha^{k_\alpha-1} \right)\bigg)\times\prod_{\beta\neq\mu,\nu} H_\beta^{(1)}\cdots H_\beta^{(k_\beta)},
\end{eqnarray}
and for $\mu=\nu$,
\begin{eqnarray}
     f_{\calO_X} \Ch\calO_{\widetilde{S_{\mu\mu}}} &=&e^{-2H_\mu}(1-e^{-H_\mu})^{k_\mu-2} \times\prod_{i_\mu=1}^{k_\mu}(1-e^{-H^{(i_\mu)}_\mu})^{m_\mu-k_\mu-1}\times\prod_{\beta\neq\mu,\nu} H_\beta^{(1)}\cdots H_\beta^{(k_\beta)}
    \nonumber\\
    &=& \left( H_\mu^{k_\mu-2}-\frac{m_\mu+1}{2}H_\mu^{k_\mu-1}+\frac{m_\mu(3m_\mu+7)}{24}H_\mu^{k_\mu}+\frac{k_\mu-m_\mu+1}{12}H_\mu^{k_\mu-2}c_2(\calS_\mu^\vee) \right)
     \nonumber\\
    &&\times \widetilde{H}_\mu^{m_\mu-k_\mu-1}\times\prod_{\beta\neq\mu} H_\beta^{(1)}\cdots H_\beta^{(k_\beta)}.
\end{eqnarray}
Notice that 
\begin{eqnarray}
    c_2(\calS_\mu^\vee)=\sum_{i<j}H_\mu^{(i)}H_\mu^{(j)}=\left\{\begin{array}{cc}
        \widetilde H_\mu & \text{if }k_\mu=2,
        \\
        0 & \text{if }k_\mu=1.
    \end{array}\right.
\end{eqnarray}

Now we can expand the integral of $Z_{\widetilde{S_{\mu\nu}}}$ in a similar way in the computation for $Z_{\widetilde{C_\mu}}$ and $Z_P$. Firstly, for any form $\omega(J)$ on $X$ one can use the adjunction formula to have (up to degree two)
\begin{eqnarray}
    \int_X\frac{\omega}{f_{\calO_X}}&=&\int_{A}\omega(H)\prod_{I=1}^K\frac{ r_I}{1-e^{- r_I}}
    \nonumber\\
    &=&\int_{A}\omega(H)\prod_{I=1}^K\left(1+\frac{ r_I}{2}+\frac{ r_I^2}{12}+\cdots\right)
    \nonumber\\
    &=&\int_{A}\omega(H)\left( 1+\frac{1}{2}c_1(\calR)+\frac{1}{12}\left(c_1^2(\calR)+c_2( \calR)\right) \right)
\end{eqnarray}
Then, the A-period of $\calO_{\widetilde{S_{\mu\nu}}}$ is given by the following expansion: for $\mu\neq\nu$, 
\begin{eqnarray}
    Z_{{\widetilde{S_{\mu\nu}}}}(\kappa)&=&\int_A \left( 1+J+\frac{1}{2}J^2+\frac{1}{24}c_2(X) \right)\left( 1+\frac{1}{2}m(H)+\frac{1}{12}(m(H)^2+c_2(\calR)) \right)
    \nonumber\\
    &&\times \prod_{\alpha=\mu,\nu} \bigg( \widetilde{H}_\alpha^{m_\alpha-k_\alpha-1}\left( -\frac{m_\alpha}{2}H_\alpha^{k_\alpha}+H_\alpha^{k_\alpha-1} \right)\bigg)\times\prod_{\beta\neq\mu,\nu} H_\beta^{(1)}\cdots H_\beta^{(k_\beta)}
    \nonumber\\
    &=&\kappa_\mu\kappa_\nu-\frac{1}{24}m_\mu m_\nu+\frac{1}{24}\int_{G(k_\mu,m_\mu)\times G(k_\nu,m_\nu)}\prod_{\alpha=\mu,\nu} \bigg( \widetilde{H}_\alpha^{m_\alpha-k_\alpha-1}H_\alpha^{k_\alpha-1}\bigg)\times c_2(\calR),
    \nonumber\\
\end{eqnarray}
and for $\mu=\nu$, 
\begin{eqnarray}
    Z_{\widetilde{S_{\mu\mu}}}(\kappa)&=&\int_A \left( 1+J+\frac{1}{2}J^2+\frac{1}{24}c_2(X) \right)\left( 1+\frac{1}{2}c_1(\calR)+\frac{1}{12}(c_1(\calR)^2+c_2(\calR)) \right)
    \nonumber\\
    &&\times\left( H_\mu^{k_\mu-2}-\frac{m_\mu+1}{2}H_\mu^{k_\mu-1}+\frac{m_\mu(3m_\mu+7)}{24}H_\mu^{k_\mu}+\frac{k_\mu-m_\mu+1}{12}H_\mu^{k_\mu-2}c_2(\calS_\mu^\vee) \right)
     \nonumber\\
    &&\times \widetilde{H}_\mu^{m_\mu-k_\mu-1}\times\prod_{\beta\neq\mu} H_\beta^{(1)}\cdots H_\beta^{(k_\beta)}
    \nonumber\\
    &=&\frac{\kappa_\mu^2}{2}-\frac{\kappa_\mu}{2}+\frac{m_\mu-m_\mu^2}{48}+\frac{1}{24}\int_{G(k_\mu,m_\mu)} \widetilde{H}_\mu^{m_\mu-k_\mu-1}H_\mu^{k_\mu-2}\times c_2(\calR)
    \nonumber\\
    &&+\frac{k_\mu-m_\mu+1}{8}\int_{G(k_\mu,m_\mu)} \widetilde{H}_\mu^{m_\mu-k_\mu-1}H_\mu^{k_\mu-2}\times c_2(\calS_\mu^\vee).
\end{eqnarray}
In particular, the last term involving integral on $G(k_\mu,m_\mu)$ equals $0$ when $k_\mu=1$ and $(3-m_\mu)/8$ when $k_\mu=2$. To further simplify the above two formulae to the last line, one should expand $c(X)=c(A)/c(\calR)$ to the second leading term and notice that only the cross term $H_\mu H_\nu$ or $H_\mu^2$ and $c_2(\calS_\mu^\vee)$ contribute in each case:
\begin{eqnarray}
    c(X)&=&\frac{\prod_{\alpha=1}^r\left( 1+m_\alpha H_\alpha+\frac{m_\alpha(m_\alpha-1)}{2}H_\alpha^2+(k_\alpha-m_\alpha+1)c_2(\calS_\alpha^\vee)+\cdots \right)}{\prod_{I=1}^K(1+r_I)}
    \nonumber\\
    &=& 1+\sum_{\alpha=1}^r\left( \frac{m_\alpha(m_\alpha-1)}{2}H_\alpha^2+(k_\alpha-m_\alpha+1)c_2(\calS_\alpha^\vee) \right)+A_2(m)-c_2(\calR)+\cdots
    \nonumber\\
\end{eqnarray}
where $A_2(m):=\sum_{\alpha<\beta} (m_\alpha m_\beta H_\alpha H_\beta)$. 

Finally, it is direct to expand the Hirzebruch-Riemann-Roch formula for $\mu\neq\nu$ as
\begin{eqnarray}
    \chi(S_{\alpha\beta},\widetilde{S_{\mu\nu}})&=&\int_A H_\alpha H_\beta  \prod_{\rho=\mu,\nu}\widetilde{H}_\rho^{m_\rho-k_\rho-1}H_\rho^{k_\rho-1}\times \prod_{\sigma\neq\mu,\nu}H_\sigma^{(1)}\cdots H_\sigma^{(k_\sigma)}
    \nonumber\\
    &=& \delta_{\alpha\mu}\delta_{\beta\nu},
\end{eqnarray}
and similarly for $\mu=\nu$ as
\begin{eqnarray}
     \chi(S_{\alpha\beta},\widetilde{S_{\mu\mu}})&=&\int_A H_\alpha H_\beta  \widetilde{H}_\mu^{m_\mu-k_\mu-1}H_\mu^{k_\mu-2}\times \prod_{\sigma\neq\mu}H_\sigma^{(1)}\cdots H_\sigma^{(k_\sigma)}
    \nonumber\\
    &=& \delta_{\alpha\mu}\delta_{\beta\mu}.
\end{eqnarray}
However, the pairing between $\widetilde{S_{00}}$ or $\widetilde{S_{0\alpha}}$ to any other cycle is zero:
\begin{eqnarray}
    \chi(\calO_{\widetilde{{S_{\alpha\beta}}}},\calB)=0,\quad \text{if }\calB\neq \calO_{S_{\rho\sigma}}.
\end{eqnarray}

\subsection{Splitting formulae} \label{sec:splitting}

Finally, we compute the topological numbers for a splitting CY4 $X$ that will be useful later. A splitting configuration CY4 $X$ is denoted as
\begin{equation}
    X=\left[\begin{array}{c|cccccc}
         \bbP^{m-1} & 1 & \cdots & 1 & 0 & \cdots & 0 
         \\
         \bbG &  n_1 & \cdots &  n_{m} &  n_{m+1} & \cdots &  n_K
    \end{array}\right]
\end{equation}
Via the conifold transition by smoothing, $X$ relates to another CY4\cite{Intriligator:2012ue} 
\begin{equation}
    X^\flat=\left[\begin{array}{c|cccccc}
         \bbG &  n &  n_{m+1} & \cdots & n_K
    \end{array}\right],\quad  n= n_1+\cdots+ n_{m}
\end{equation}
For simplicity, we denote the ambient space as
\begin{equation}
    Y:=\left[\begin{array}{c|cccccc}
         \bbG &  n_{m+1} & \cdots &  n_K
    \end{array}\right],
\end{equation}
then $X^\flat$ is a hypersurface in $Y$ and $X$ is a complete intersection in $\bbP^{m-1}\times Y$ given by line bundles $\calL_1\boxtimes\calO_{\bbP^{m-1}}(1),\cdots,\calL_{m}\boxtimes\calO_{\bbP^{m-1}}(1)$. Denote $L$ as the hyperplane class of $\bbP^{m-1}$, $H$ as that of $Y$ (e.g. $c_1(\calL_I)=n_I$), the fundamental class of $X$ can be expanded as
\begin{equation}
    [X]=\prod_{I=1}^{m} (L+c_1(\calL_I))=L^{m-1} A_1(\calL_I)+L^{m-2}A_2(\calL_I)+\cdots
\end{equation}
where
\begin{equation}
    A_k(\calL_I)=\sum_{I_1<\cdots< I_k} c_1(\calL_{I_1})\cdots c_1(\calL_{I_k}),\quad A_1(\calL_I)=\sum_{I=1}^{m}c_1(\calL_I)=c_1(Y)
\end{equation}
Then for any form $\omega(L,H)=\omega_0(H)+L\omega_1(H)+\cdots$,
\begin{equation}
\begin{aligned}
    \int_{X}\omega(L,H) =&\int_{\bbP^{m-1}\times Y} \omega (L,H)[X]
    \\
    =&\int_{Y} \omega_0A_1+\omega_1 A_2+\omega_2 A_3+\cdots
    \\
    =&\int_{X^\flat} \omega_0+\int_Y \omega_1 A_2+\cdots.
\end{aligned}
\end{equation}
Thus the topological numbers for $X$ that only involve $Y$ directions are the same as the numbers for $X^\flat$, while other numbers that involve $L$ can be expanded as the integral over $Y$ using $A_k(\calL_I)$.

The Chern class of $X^\flat$ in Fano 5-fold $Y$ is
\begin{eqnarray}
    c(X^\flat)&=&\frac{c(Y)}{c(-K_Y)}
    \nonumber\\
    &=&1+c_2(Y)+c_3(Y)-c_1(Y)c_2(Y)
    \nonumber\\
    &+&c_4(Y)-c_1(Y)c_3(Y)+c_1(Y)^2c_2(Y)
\end{eqnarray}
Similarly, expand the Chern class of $X$ in $\bbP^{m-1}\times Y$ one has
\begin{eqnarray}
    c(X)&=&\frac{c(Y)c(\bbP^{m-1})}{c(\calN\boxtimes\calO_{\bbP^{m-1}}(1))}
    \nonumber\\
    &=&c(X^\flat)-A_2( n)+A_1( n)A_2( n)-A_3( n)
    \nonumber\\
    &+&A_2( n)^2+A_1( n)A_3( n)-A_1( n)^2A_2( n)-A_4( n)-A_2( n) c_2(Y)
   \nonumber\\
   &+&L \left(A_1( n) +2 A_2( n)-A_1( n)^2+3
   A_3( n)-4 A_1( n)A_2( n)+A_1( n)^3+A_1( n) c_2(Y)\right)
   \nonumber\\
   &+&L^2 \left(-A_1( n)+2 A_1( n)^2-3 A_2( n)\right)+L^3 A_1( n)
\end{eqnarray}
In where:
\begin{equation}
    c_k(\calN\boxtimes\calO_{\bbP^{m-1}}(1))=\sum_{j=0}^k\begin{pmatrix}
        m-j \\ k-j
    \end{pmatrix} L^{k-j}c_j(\calN).
\end{equation}

For any form
\begin{eqnarray}
    \omega=\sum_{i=0}^4L^i\omega_i(H),
\end{eqnarray}
expanding the induced fundamental class gives
\begin{eqnarray}
    \int_X \omega &=&\int_{\bbP^{m-1}\times Y}\omega  \prod_{I=1}^{m}\left( L+ n_I\right)
    \nonumber\\
    &=&\int_{\bbP^{m-1}\times Y} \omega\left( L^{m-1}A_1(n)+L^{m-2}A_2(n)+\cdots \right)
    \nonumber\\
    &=&\int_{Y}\sum_{i=0}^4 A_{i+1}(n)\omega_i.
\end{eqnarray}
One has
\begin{eqnarray}
    c_{0000}&=& \int_Y A_5( n)
    \\
    c_{000\alpha}&=&\int_Y A_4( n) H_\alpha
    \\
    c_{00\alpha\beta}&=&\int_Y A_3( n)H_\alpha H_\beta
    \\
    c_{0\alpha\beta\gamma}&=&\int_Y A_2( n)H_\alpha H_\beta H_\gamma
    \\
    c_{\alpha\beta\gamma\omega}&=&c_{\alpha\beta\gamma\omega}(X^\flat)
\end{eqnarray}
\begin{eqnarray}
    c_{00}&=&\frac{1}{24}\int_Y c_2(Y)A_3( n)+A_1( n)A_4( n)-A_2( n)A_3( n)
    \\
    c_{0\alpha}&=&\frac{1}{24}\int_Y \big( c_2(Y)A_2( n)+A_1( n)A_3( n)-A_2( n)^2 \big) H_\alpha
    \\
    c_{\alpha\beta}&=&c_{\alpha\beta}(X^\flat)
\end{eqnarray}
\begin{eqnarray}
    c_0&=&\frac{\zeta(3)}{(2\pi i)^3}\bigg(\int_{X^\flat}( A_2^2-A_1A_3 ) +\int_Y( A_2A_3-A_1A_4 )\bigg),
    \\
    c_{\alpha}&=&c_{\alpha}(X^\flat)+\frac{\zeta(3)}{(2\pi i)^3}\int_Y 2(A_2^2-A_1A_3)H_\alpha.
\end{eqnarray}
Notice that different to the CY3 case, $c_\alpha$ is changed under splitting. For simplicity, we only need to compute the change of Euler number as\cite{Brunner:1996bu,Intriligator:2012ue}
\begin{eqnarray}
    \chi(X)&=&\chi(X^\flat)+3\int_{X^\flat} A_1A_3-A_2^2
\end{eqnarray}
All the results are well-organized into two classes about the partition $n_I$ as:
\begin{equation}
    N_4( n)=A_2^2-A_1A_3,\quad N_5( n)=A_2A_3-A_1A_4.
\end{equation}

Finally, the classical A-period of $\widetilde{S_{00}}$ and $\widetilde{S_{0\alpha}}$ specialized to determinantal CY4 is as
\begin{eqnarray}
    Z_{\widetilde{S_{0\alpha}}} &=&\kappa_0\kappa_\alpha-\frac{1}{24}n^{(\alpha)},
    \\
     Z_{\widetilde{S_{00}}}&=&\frac{\kappa_0^2}{2}-\frac{\kappa_0}{2}.
\end{eqnarray}

\section{Universal monodromy from GLSM} \label{sec:monodromy}

Now we will fix the period basis for a splitting CY4 $X$ as
\begin{eqnarray}
   \vec\Pi(\kappa)= \langle Z_{\calO_X},\ Z_{D_0},\ Z_{D_{\alpha}},\ Z_{{S_{00}}},\ Z_{{S_{0\alpha}}},\ Z_{{S_{\alpha\beta}}},\ Z_{S_\alpha},\ Z_{\widetilde{C_0}},\ Z_{\widetilde{C_\alpha}},\ Z_P \rangle
\end{eqnarray}
Notice that when $m=2$, $\calO_{S_{00}}$ class is empty and can be omitted. The monodromy action on this basis can be computed using grade restriction rule and window shift as in \cite{Lin:2026icu}. In particular, the results there are universal on the chain complex level and thus can be directly recycled for CY4 cases:
\begin{thm}{(Section 3.2 of \cite{Lin:2026icu})}\label{thm:monodromyCY3}
    Given a GLSM that realizes a splitting CY n-fold $X_n\subset \bbP^{m-1}\times Y_{n+1}$, the abelian window shift action $M$ is isomorphic to the action $L_{K_Y}$ when it acts on the B-branes corresponding to $\calO_X,\ \calO_{D_\alpha},\ \calO_{\widetilde{C_\alpha}},\ \calO_P$.\footnote{The action on later two objects is derived at the central charge level.} The action on $\calO_{D_0}$ is numerically equivalent to $L_{K_Y}\left(\calO_{D_0}\oplus\calC[1]\right)$ where
    \begin{eqnarray}
        \calC:=\bigg(\mathcal{O}^{\oplus m}_{X}\rightarrow \mathcal{O}_{X}(-1,0)\oplus\bigoplus_{I=1}^{m}\mathcal{O}_{X}(0,{n}_{I})\rightarrow \mathcal{O}_{X}(-1, n)\bigg).
    \end{eqnarray}
\end{thm}
We will evaluate the A-period of $\calC$ in CY4. In addition to the above results, the action of $M$ on $\calO_{S_{0\alpha}}$ and $\calO_{S_{\alpha\beta}}$ can also be directly induced from $M(\calO_X)$ and $M(\calO_{D_0})$ using their defining exact sequences. As a consequence, we will only need to compute $M(\calO_{S_{00}})$ for $X$, which will be derived in this section at the brane factor level for simplicity.

The computation of each monodromy is conducted later in this section. In summary, we have
\begin{thm} \label{thm:monodromy}
The window shift monodromy in a CY4 flop is given by the following action on charge lattice $\vec\Pi$:
\begin{eqnarray}
    M=T\cdot L_{K_Y},
\end{eqnarray}
in where $L_{K_Y}:=\prod_{\alpha=1}^rL_\alpha^{-n^{(\alpha)}}$ for each large volume monodromy $L_\mu$ given by shifting $\kappa_\alpha\mapsto \kappa_\alpha+\delta_{\mu\alpha}$ as
\begin{eqnarray}
    \vec \Pi(\kappa)\cdot L_\mu=\vec \Pi(\kappa+\mathbf e_{\mu}),
\end{eqnarray}
which is fully determined by the classical part of A-periods in terms of their topological numbers:
\begin{eqnarray}\label{eqn:L}
    \begin{pmatrix}
        Z_{X}
    \\
    Z_{D_\alpha}
    \\
    Z_{S_{\alpha\beta}}
    \\
    Z_{S_\alpha}
    \\
    Z_{\widetilde{C_\alpha}}
    \\
    Z_P
    \end{pmatrix}^t (\kappa+\mathbf{e}_\mu) &=& \begin{pmatrix}
       Z_X+Z_{D_\mu}+Z_{S_{\mu\mu}}+c_{\mu\mu\mu\alpha} Z_{\widetilde{C_\alpha}}-\frac{c_{\mu\mu\mu\mu}}{2}Z_P
    \\
    Z_{D_\alpha}+Z_{S_{\mu\alpha}}+c_{\mu\mu\alpha\beta}Z_{\widetilde{C_{\beta}}}+\frac{c_{\mu\mu\alpha\alpha}}{2}Z_P
    \\
    Z_{S_{\alpha\beta}}+c_{\alpha\beta\mu\rho} Z_{\widetilde{C_\rho}}+b_{\alpha\beta\mu}Z_P
    \\
    Z_{S_\alpha}+\widetilde{c_{\alpha\mu\beta}}Z_{\widetilde{C_\beta}}+\frac{\widetilde{c_{\alpha\mu\mu}}-\widetilde{c_{\alpha\alpha\mu}}}{2}Z_P
    \\
    Z_{\widetilde{C_\alpha}}+\delta_{\mu\alpha}Z_P
    \\
    Z_P
    \end{pmatrix}^t
\end{eqnarray}
for
\begin{eqnarray}
    b_{\alpha\beta\mu}&=&\frac{1}{2}(c_{\alpha\beta\mu\mu}-c_{\alpha\alpha\beta\mu}-c_{\alpha\beta\beta\mu}).
\end{eqnarray}
And the action of $T$ acts as 
\begin{eqnarray} \label{eqn:T}
   \begin{pmatrix}
        Z_{D_0^{\;}}
        \\ 
        Z_{S_{00}^{\;}}
        \\
        Z_{S_{0\alpha}^{\;}} 
    \end{pmatrix}^t\cdot T=
    \begin{pmatrix}
        Z_{D_0}-N_5\cdot Z_{\widetilde{S_{00}}}-N_4^{(\beta)}Z_{\widetilde{S_{0\beta}}}-(N_4/2) Z_{\widetilde{C_0}}
        \\
        Z_{S_{00}}-N_5Z_{\widetilde{C_0}}
        \\
        Z_{S_{0\alpha}}-N_4^{(\alpha)}Z_{\widetilde{C_0}}
    \end{pmatrix}^t
\end{eqnarray}
\end{thm}

Using the pairing matrix in section \ref{sec:period}, $T$ can be written as the following form, which is an EZ-twist for a surface class $S$ collapsing to a genus $g$ curve with generic fiber $F$ \cite{Aspinwall:2001zq,Cota:2019cjx}: \footnote{It is crucial for $S$ being an exceptional surface with $F\cong\bbP^1$ fiber and any invertible sheaf on $S$ is EZ-spherical, such that the original CY3 formula for divisor degeneration is applicable for surface degeneration in CY4 \cite{Horja:2001cp, donovan2024derived}.}
\begin{eqnarray}
    Z_{ T(\calB)} &=& Z_\calB-\chi(\calO_S+(1-g)\calO_F,\calB)Z_{\calO_F}+\chi(\calO_F,\calB)Z_{\calO_S},
\end{eqnarray}
in where $1-g=-N_4/2$ and
\begin{eqnarray}
    Z_{\calO_S} &=& Z_\calC+(1-g) Z_{\calO_F},
    \nonumber\\
    &=& N_5Z_{\calO_{\widetilde{S_{00}}}}+\sum_\alpha N_4^{(\alpha)}Z_{\calO_{\widetilde{S_{0\alpha}}}}
    \\
    Z_{\calO_F} &=&Z_{\calO_{\widetilde{C_0}}}.
\end{eqnarray}
Later in section \ref{sec:example}, we will also illustrate that $T$ can also be decomposed into braid group actions of $L_\mu$ and some spherical twist $T_{E_i}$ of objects $E_i$ with support on $X$ \cite{Aspinwall:2001zq,halpern2016autoequivalences,donovan2024derived,Lin:2026icu}. The spherical twist $T_E$ on our basis is given by \cite{Seidel:2000ia,Aspinwall:2001zq}
\begin{eqnarray}
   Z_{ T_{E}(\calB) }:=Z_\calB-\chi(E,\calB)Z_{E}.
\end{eqnarray}
And the most important one is the spherical twist of structure sheaf $\calO_X$:
\begin{eqnarray}
    T_{\calO_X}&=&\Id-\begin{pmatrix}
        2 & -l_\alpha & s_{\alpha\beta} & \widetilde{s_\alpha} &  \ \ 0 & \ \ 1
        \\
         & 0
         \\
         && 0
         \\
         &&& 0
         \\
         &&&&  \ \ 0
         \\
         &&&&& \ \ 0
    \end{pmatrix},
    \\
   l_\alpha&=&\frac{1}{24}c_{\alpha\alpha\alpha\alpha}+c_{\alpha\alpha},
   \\
   s_{\alpha\beta}&=&\frac{1}{6}c_{\alpha\beta\beta\beta}+\frac{1}{4}c_{\alpha\alpha\beta\beta}+\frac{1}{6}c_{\alpha\alpha\alpha\beta}+2c_{\alpha\beta}, 
   \\
   \widetilde{s_\alpha} &=& \frac{1}{6}\widetilde{c_{\alpha\alpha\alpha}}-\widetilde{c_{\alpha\alpha}}+2\widetilde{c_\alpha}
\end{eqnarray}
We will determine another spherical twist $T_{\calS_X}$ in our examples.

\subsection*{Divisor class $D_0$}

The monodromy action $M(\calO_{D_0})$ can be directly given by evaluating $Z_{\calC}$ as in CY3 cases. The object $\calC$ has Chern character (up to the fourth degree)
\begin{eqnarray}
    \Ch\calC&=&\frac{f_{\calC}(H/2\pi)}{f_{\calE_-}(H/2\pi)}=\sum_{I=1}^{m}\left(1-e^{ n_I(H)}\right)-e^{-L}\left(1-e^{ n(H)}\right)
    \nonumber\\
    &=&A_2-A_1L+\frac{1}{2}LA_1\left(L-A_1\right)-\frac{1}{2}(A_3-A_1A_2)
    \nonumber\\
&-&\frac{1}{6}A_1^3L+A_1^2\left(\frac{1}{4}L^2+\frac{1}{6}A_2\right)-\frac{1}{6}A_1(L^3+A_3)-\frac{1}{12}A_2^2+\frac{1}{6}A_4
\end{eqnarray}
Now $Z_\calC$ is evaluated by
\begin{eqnarray}
    Z_\calC=\int_X \left( 1+\kappa_0L+ \kappa(H)+\frac{1}{2}\kappa_0^2L^2+\frac{1}{2}\kappa^2+\kappa_0(L)\kappa \right)\left( 1+\frac{c_2(X)}{24} \right)\Ch\calC.
\end{eqnarray}
The top degree is evaluated as a similar Thom-Porteous formula on points
\begin{eqnarray}
        &&\int_X-\frac{1}{6}A_1^3L+A_1^2\left(\frac{1}{4}L^2+\frac{1}{6}A_2\right)-\frac{1}{6}A_1(L^3+A_3)-\frac{1}{12}A_2^2+\frac{1}{6}A_4
        \nonumber\\
        &=&\frac{1}{12}\int_{Y}A_1(A_1A_3-A_2^2)
        \nonumber\\
        &=&-\frac{1}{12}\int_{X^\flat} N_4(n)
\end{eqnarray}
The third degree is evaluated as
\begin{equation}
\begin{aligned}
    &\frac{1}{2}\int_X \left(\kappa(H)+\kappa_0L \right)\left( A_1L^2+A_1A_2-A_1^2L-A_3 \right)
    \\
    =&\frac{1}{2}\int_Y \kappa(H) \left( A_1A_3+A_1^2A_2-A_1^2A_2-A_1A_3 \right)
    \\
    +&\frac{1}{2}\int_Y \kappa_0\left( A_1A_4+A_1A_2^2-A_1^2A_3-A_2A_3 \right)
    \\
    =&\frac{\kappa_0}{2}\left(\int_{X^\flat}N_4(n)-\int_Y N_5(n)\right)
\end{aligned}
\end{equation}

And finally, the second degree:
\begin{eqnarray}
    &&\int_X  \left( \frac{1}{2}\kappa(H)^2+\frac{1}{2}\kappa_0^2L^2+\kappa_0\kappa(H)L + \frac{c_2(X)}{24} \right) (A_2-A_1L)
    \nonumber\\
    &=&\frac{\kappa_0^2}{2}\int_Y N_5( n)+\kappa_0\kappa_\alpha\int_{Y}H_\alpha N_4 +\frac{1}{24}\int_{X^\flat} N_4( n) 
\end{eqnarray}
In conclusion, the contraction class $\calC$ in the $T_N$ action on CY4 is a class in surfaces and curves with
\begin{eqnarray}
    Z_\calC&=&N_5\cdot Z_{\widetilde{S_{00}}}+N_4^{(\alpha)}\cdot Z_{\widetilde{S_{0\alpha}}}+\frac{N_4}{2}\cdot Z_{\widetilde{C_0}}
\end{eqnarray}
where
\begin{eqnarray}
    N_4^{(\alpha)}=\int_{Y}H_\alpha N_4(n).
\end{eqnarray}

\subsection*{Surface classes $S_{0\alpha}$, $S_{\alpha\beta}$}

It is direct to read the result for these surface classes by applying the result on $\calO_X$ and $\calO_{D_0}$. Note that monodromy $M$ on  satisfies $M(\calO_X( \vec q))=(L_{K_Y}(\calO_X))(\vec q)$, the surface classes from exact sequence ($\alpha,\beta\neq0$)
\begin{equation}
    0\rightarrow\calO_X(-H_\alpha-H_\beta)\rightarrow \calO_X(-H_\alpha)\oplus\calO_X(-H_\beta)\rightarrow\calO_X\rightarrow\calO_{S_{\alpha\beta}}\rightarrow0
\end{equation}
and
\begin{eqnarray}
     0\rightarrow\wedge^2\calS_\alpha|_X\rightarrow \calS_\alpha|_X\rightarrow\calO_X\rightarrow\calO_{S_{\alpha}}\rightarrow0
\end{eqnarray}
admit monodromy actions on each term in the complex, and the result is
\begin{eqnarray}
    M(\calB)=L_{K_Y}(\calB),\qquad \text{for  }\calB=\calO_{S_{\alpha\beta}},\ \calO_{S_\alpha}.
\end{eqnarray}

Similarly, the monodromy on $\calO_{S_{0\alpha}}$ for $\alpha\neq0$ can be reduced from the monodromy on $\calO_{D_0}$ by the exact sequence
\begin{eqnarray}
    0\rightarrow \calO_{D_0}(-J_\alpha)\rightarrow \calO_{D_0}\rightarrow\calO_{S_{0\alpha}}\rightarrow0.
\end{eqnarray}
Thus the result is
\begin{eqnarray}
    M(\calO_{S_{0\alpha}})=  L_{K_Y}(\Cone(\calC_\alpha\rightarrow\calO_{S_{0\alpha}}))
\end{eqnarray}
where
\begin{equation}
    0\rightarrow\calC(-J_\alpha)\rightarrow \calC\rightarrow \calC_\alpha\rightarrow0.
\end{equation}
Its Chern character is given by (up to degree four)
\begin{eqnarray}
    \Ch\calC_\alpha&=&\left(\sum_{I=1}^{m}\left(1-e^{ n_I}\right)-e^{-L}\left(1-e^{ n}\right)\right)\left( 1-e^{-H_\alpha} \right)
    \nonumber\\
   &=& \frac{1}{2}\left( A_1L-A_2 \right)H_\alpha^2+\frac{1}{2}\left( A_1A_2-A_3+A_1L^2-A_1^2L \right)H_\alpha
    \nonumber\\
    &+& (A_2-A_1 L)H_\alpha
\end{eqnarray}
and finally, its central charge is (note that the degree four part totally disappears by splitting formulae)
\begin{eqnarray}
    Z_{\calC_\alpha}&=&\int_X \left( 1+\kappa_0L+\vec \kappa(H) \right)\Ch\calC_\alpha
    \nonumber\\
    &=& \kappa_0\int_Y H_\alpha N_4 \equiv N_4^{(\alpha)}\cdot Z_{\widetilde {C_0}}.
\end{eqnarray}

\subsection*{Surface class $S_{00}$}

Comparing to the existing results in CY3 cases, the only class in CY4 that needs to compute from strach is the surface class $S_{00}$, which needs more grade restrictions. Since we focus on A-periods, here we show a simplified computation on brane factors of grade restriction and window shift, following exactly the same recipe in \cite{Lin:2026icu}. We assume again that $n_I(H)=n_I^{(\alpha,i_\alpha)}H_\alpha^{(i_\alpha)}$ is the Chern root of $\calN$.

The brane factor of $S_{00}$ is denoted as
\begin{eqnarray}
    f_{\calO_{S_{00}}}=\left( 1-q_0^{-1} \right)^2f_{\calO_X}=\left( 1-2q_0^{-1}+q_0^{-2} \right)\times\prod_{I=1}^m(1-q_0^{-1} {q}^{-1}_I)f_{\calO_Y}
\end{eqnarray}
where
\begin{eqnarray}
    \vec q^{d}:=\sum_{\alpha,i_\alpha}q_{\alpha,i_\alpha}^{d^{(\alpha,i_\alpha)}}=\exp\left(2\pi \sum_\alpha d^{(\alpha,i_\alpha)}\sigma^{(i_\alpha)}_\alpha\right).
\end{eqnarray}
and
\begin{eqnarray}
    q_I:=\vec{q}^{\ n_I},\qquad q^n:=q_1\cdots q_m.
\end{eqnarray}
where $f_{\calO_X}$ is
\begin{eqnarray}
    f_{\calO_X}&=&\prod_{I=1}^m\left( 1-q_0^{-1}q_I^{-1} \right)f_{\calO_Y}
    \nonumber\\
    &=&\left( 1+q_0^{-1}\sum_I(- q_I^{-1}) +q_0^{-2}\sum_{I<J}(-q_I^{-1})(-q_J^{-1})-\cdots+(-1)^m q_0^{-m} q^{-n} \right)f_{\calO_Y}
    \nonumber\\
    &=:&\bigg( 1+q_0^{-1} A_1(-q_I^{-1}) +\cdots
    \nonumber\\
    &&+ q_0^{-m+2}A_{m-2}(-q_I^{-1})+ q_0^{-m+1}A_{m-1}(-q_I^{-1})+ q_0^{-m}A_{m}(-q_I^{-1}) \bigg)f_{\calO_Y}
    \nonumber\\
\end{eqnarray}
On the brane factor level, the grade restriction rule of object $\calW(-m,d)$, $\calW(-m-1,d)$ and $\calW(-m-2,d)$ are given by (with underline terms as empty brane factors)
\begin{eqnarray}
    q_0^{-m}\vec q^d&\cong& \left( q_0^{-m}-\underline{(q_0^{-1}-1)^m} \right)\vec q^d
    \nonumber\\
    &=& -\left(m(-1) q_0^{-m+1}+\begin{pmatrix}
        m \\ 2
    \end{pmatrix}(-1)^2q_0^{-m+2}+\cdots +(-1)^m\right) \vec q^d
\end{eqnarray}
Then applying $q_0^{-1}$, one has
\begin{eqnarray}
     q_0^{-m-1} \vec q^d &\cong& q_0^{-1}\left( q_0^{-m}-\underline{(q_0^{-1}-1)^m} \right)\vec q^d
    \nonumber\\
    &=&\left( mq_0^{-m}-\begin{pmatrix}
        m\\2
    \end{pmatrix}q_0^{-m+1}+\cdots \right) \vec q^d
    \nonumber\\
    &\cong& \bigg( mq_0^{-m}-\underline{m(q_0^{-1}-1)^m} -\begin{pmatrix}
        m\\2
    \end{pmatrix}q_0^{-m+1}+\cdots \bigg)\vec q^d
    \nonumber\\
    &=&\bigg(  \begin{pmatrix}
        m+1\\2
    \end{pmatrix}q_0^{-m+1}+\cdots \bigg)\vec q^d
\end{eqnarray}
thus the GRR result is
\begin{eqnarray}
   q_0^{-m-1}&\cong&\bigg( q_0^{-m-1} - q_0^{-1}(q_0^{-1}-1)^m-m(q_0^{-1}-1)^m  \bigg)\vec q^d.
\end{eqnarray}
Finally, applying $q_0^{-1}$ again, one has
\begin{eqnarray}
    q_0^{-m-2}\vec q^d &\cong& q_0^{-1}\bigg( q_0^{-m-1} - q_0^{-1}(q_0^{-1}-1)^m-m(q_0^{-1}-1)^m  \bigg)\vec q^d
    \nonumber\\
    &\cong&\left(  \begin{pmatrix}
        m+1\\2
    \end{pmatrix}q_0^{-m}-\underline{\begin{pmatrix}
        m+1\\2
    \end{pmatrix}(q_0^{-1}-1)^m}+\cdots \right)\vec q^d
\end{eqnarray}
and thus the GRR result is
\begin{eqnarray}
    q_0^{-m-2}\vec q^d &\cong&\left(q_0^{-m-2}-(q_0^{-1}-1)^m\left(   q_0^{-2}+mq_0^{-1}+\frac{m(m+1)}{2} \right)\right)\vec q^d
\end{eqnarray}

Next, we compute the monodromy action on each grade restricted brane component. Any such action is essentially given by the action on $\calW(0,d)$ as
\begin{eqnarray}
    \vec q^d\mapsto \vec q^d \left(1- {\prod_{I=1}^m\left(1-q_0^{-1}q_I^{-1}\right)}\right)=:\vec q^d  \left(1- f_{\calE_-}\right)
\end{eqnarray}
in where $f_{\calE_-}$ is the empty brane factor of negative phase. Then it is sufficient to apply the above action on grde restricted $f_{\calO_{S_{00}}}$ by only applying it on the constant term of $q_0$ in $f_{\calO_{S_{00}}}$. For instance, the monodromy action of $q_0^{-m}\vec q^d$ is
\begin{eqnarray}
    q_0^{-m}\vec q^d&=& \left( q_0^{-m}-{(q_0^{-1}-1)^m} \right)\vec q^d
    \nonumber\\
    &=&\left(-m(-1)q_0^{-m+1} \cdots-m(-1)^{m-1}q_0^{-1}-(-1)^m \right)\vec q^d
    \nonumber\\
    &\mapsto& q_0^{-m}\vec q^d +(-1)^mf_{\calE_-}\vec q^d
\end{eqnarray}
and
\begin{eqnarray}
   q_0^{-m-1}{\vec q}^{\ d} &\cong&\bigg( q_0^{-m-1} - q_0^{-1}(q_0^{-1}-1)^m-m(q_0^{-1}-1)^m  \bigg)\vec q^d
   \nonumber\\
   &=&\bigg( \cdots-m(-1)^m \bigg)\vec q^d
   \nonumber\\
   &\mapsto& q_0^{-m-1}\vec q^d +m(-1)^mf_{\calE_-}\vec q^d
\end{eqnarray}
and similarly
\begin{eqnarray}
    q_0^{-m-2}\vec q^d &\cong&\left(q_0^{-m-2}-(q_0^{-1}-1)^m\left(   q_0^{-2}+mq_0^{-1}+\frac{m(m+1)}{2} \right)\right)\vec q^d
    \nonumber\\
    &=&\left( \cdots -\frac{m(m+1)}{2}(-1)^m \right)\vec q^d
    \nonumber\\
    &\mapsto&  q_0^{-m-2}\vec q^d +(-1)^m\frac{m(m+1)}{2}f_{\calE_-}\vec q^d
\end{eqnarray}

Thus, the final result is simply given by applying monodromy action on the following terms (Note that $A_{m-k}(-q_I^{-1})=(-1)^{m-k}A_k(q_I)q^{-n}$)
\begin{eqnarray}
    f_{\calO_{S_{00}}}&=&\left( 1-q_0^{-1} \right)^2\bigg( 1+q_0^{-1} A_1(-q_I^{-1}) +\cdots
    \nonumber\\
    &&+ q_0^{-m+2}A_{m-2}(-q_I^{-1})+q_0^{-m+1}A_{m-1}(-q_I^{-1})+ q_0^{-m}A_{m}(-q_I^{-1}) \bigg)f_{\calO_Y}
    \nonumber\\
    &=& \left((-1)^m q^n+\cdots +(A_{2}(q_I)+2A_{1}(q_I)+1)q_0^{-m}-(A_{1}(q_I)+2)q_0^{-m-1}+ q_0^{-m-2}  \right)(-1)^m q^{-n} f_{\calO_Y}
    \nonumber\\
    &\mapsto&f_{\calO_{S_{00}}}-\left( q^n-A_2+(m-2)A_1 - \begin{pmatrix}
        m-1\\2
    \end{pmatrix} \right)f_{\calO_X}q^{-n}
\end{eqnarray}
Moreover, it can be written as the following form that agrees with the monodromy action with $L_{K_Y}$:
\begin{eqnarray}
    f_{M(\calO_{S_{00}})}=f_{\calO_{S_{00}}} q^{-n}-f_{\calC'}
\end{eqnarray}
where
\begin{eqnarray}
    f_{\calC'}=\left((1-q_0^{-1})^2(1-q^n) +q^n- A_{2}(q_I)+(m-2)A_{1}(q_I)-\frac{(m-1)(m-2)}{2}\right)f_{\calO_X} q^{-n}.
    \nonumber\\
\end{eqnarray}

Finally, the Chern character of component $\calC'$ is given by (up to degree four)
\begin{eqnarray}
    \Ch\calC'&=&\left( 1-e^{-L} \right)^2 \left( 1 - e^{n} \right)-e^{n}-\sum_{I<J} e^{n_I+n_J}+(m-2)\sum_{I=1}^m e^{n_I} - \begin{pmatrix}
        m-1\\2
    \end{pmatrix}
    \nonumber\\
    &=& A_3(n)-L^2A_1(n)+L^3 A_1(n)-A_4(n)+\frac{1}{2}\left( L^2A_1(n)^2 - A_1(n)A_3(n) \right)
\end{eqnarray}
Using
\begin{eqnarray}
    \sum_{I=1}^m e^{n_I}&=&m+A_1(n)+\frac{1}{2}\left( A_1(n)^2-2A_2(n) \right)
    \nonumber\\
    &&+\frac{1}{6}\left( A_1(n)^3-3A_1(n)A_2(n)+3A_3(n) \right)
    \nonumber\\
    &&+\frac{1}{24}\left( A_1^4-4A_1^2A_2+4A_1 A_3+2A_2^2-4A_4 \right)
\end{eqnarray}
and its central charge is evaluated as
\begin{eqnarray}
    Z_{\calC'}&=& \int_X \left( 1+\kappa_0 L+\kappa(H) \right)\Ch\calC'
    \nonumber\\
    &=& \int_Y \kappa_0\left( A_2A_3-A_1A_4 \right) 
    \nonumber\\
    &=& N_5 Z_{\widetilde{C_0}}.
\end{eqnarray}

\section{Examples}\label{sec:example}

Here we illustrate the decomposition of $M$ into spherical twists on two families of examples: splitting configurations in $\bbP^5[6]$ and $G(2,5)[4,1]$.

\subsection{$\bbP^5[6]$}

The topological data of $\bbP^5[6]$ is computed as
\begin{eqnarray}
    c_{1111}=6,\ c_{11}=\frac{15}{4},\ c_1=(-420)\frac{\zeta(3)}{(2\pi i)^3},\ \chi(X)=2610.
\end{eqnarray}
We start with the fundamental basis
\begin{eqnarray}
    \vec\Pi=\langle \calO_X,\calO_{D_0},\calO_{D_1},\calO_{{S_{00}}},\calO_{{S_{01}}},\calO_{S_{11}},\calO_{\widetilde{C_0}},\calO_{\widetilde{C_1}},\calO_P \rangle.
\end{eqnarray}
Then the ansatz for matrices are
\begin{eqnarray}
    L_1&=&\left(
\begin{array}{ccccccccc}
 1 &  &  &  &  &  &  &  &  \\
 0 & 1 &  &  &  &  &  &  &  \\
 1 & 0 & 1 &  & &  &  &  &  \\
 0 & 0 & 0 & 1 &  &  &  &  &  \\
 0 & 1 & 0 & 0 & 1 &  &  &  &  \\
 1 & 0 & 1 & 0 & 0 & 1 &  &  &  \\
 c_{0111} & c_{0011} & c_{0111} & c_{0001} & c_{0011} & c_{0111} &\ 1 &  &  
 \\
 6 & c_{0111} & 6 & c_{0011} & c_{0111} & 6 & \ 0 &\ \ 1 &  
 \\
 -3 & -\frac{c_{0011}}{2} & -3 & \frac{c_{0011}}{2}-c_{0001} & -\frac{c_{0011}}{2} & -3
   & \ 0 & \ \ 1 &\ \ 1 
\end{array}
\right)
\end{eqnarray}
and
\begin{eqnarray}
    T_{\calO_X}=\begin{pmatrix}
        -1 & \ \ l_0 & \ \ \ \ 4 & \ -s_{00} & -s_{01} & \  -11 & \ \  0 & \ \ \ 0 & \ \  -1
        \\
         & \ \ 1
         \\
         && \ \ \ \  1
         \\
         &&&\ \ \ 1
         \\
         &&&&\  1
         \\
         &&&&&\ 1
         \\
         &&&&&& \ \ 1
         \\
         &&&&&&& \ \ \ \ 1
         \\
         &&&&&&&& \ \ 1
    \end{pmatrix}
\end{eqnarray}
It is a direct computation to show that $(T_{\calO_X}L_1)^6$ only acts as the following while keeping other classes unchanged:
\begin{eqnarray}
    Z_{D_0}\cdot (T_{\calO_X}L_1)^6 &=&  Z_{D_0}+ 6 Z_{S_{01}}-c_{0111}Z_{S_{11}}+\left( -4c_{0111}^2 + \frac{1}{2}c_{0011}c_{0111}+21c_{0011} \right) Z_{_{\widetilde{C_0}}}
    \nonumber\\
    &&+3(c_{0011}-c_{0111})Z_{\widetilde{C_1}}+\left( 10c_{0111}-12 c_{01}-\frac{3}{2}c_{0011}-c_{0001} \right) Z_P
    \\
    Z_{S_{00}}\cdot (T_{\calO_X}L_1)^6 &=& Z_{S_{00}}-\left( c_{0011}c_{0111}-6c_{0001} \right) Z_{\widetilde{C_0}}
    \\
    Z_{S_{01}}\cdot (T_{\calO_X}L_1)^6 &=& Z_{S_{01}}-\left( c_{0111}^2-6c_{0011} \right) Z_{\widetilde{C_1}}
\end{eqnarray}
It is direct to derive (note that $A_k(n)$ is proportional to $H^k$)
\begin{eqnarray}
     c_{0011}c_{0111}-6c_{0001} &=& \int_{\bbP^5} A_3(n) H^2\int_{\bbP^5} A_2(n) H^3- \int_{\bbP^5} A_1(n)H^4 \int_{\bbP^5} A_4(n)H
     \nonumber\\
     &=& A_2(n)A_3(n)-A_1(n)A_4(n)
     \nonumber\\
     &\equiv& \int_{\bbP^5} N_5(n)
     \\
      c_{0111}^2-6c_{0011}&=& A_2(n)^2-A_1(n)A_3(n)
      \nonumber\\
      &\equiv& \int_{X^\flat} N_4(n)
\end{eqnarray}
However, it is quite non-trivial to show that the combination in $Z_{D_0}$ gives $Z_{\widetilde{S}_{00}}$ and $Z_{\widetilde{S_{01}}}$ with desired coefficients:
\begin{eqnarray}
    Z_{D_0}\cdot (T_{\calO_X}L_1)^6 &=&6 Z_{S_{01}}-c_{0111}Z_{S_{11}}+\left( -4c_{0111}^2 + \frac{1}{2}c_{0011}c_{0111}+21c_{0011} \right) Z_{_{\widetilde{C_0}}}
    \nonumber\\
    && + 3(c_{0011}-c_{0111})Z_{\widetilde{C_1}}+\left( 10c_{0111}-12 c_{01}-\frac{3}{2}c_{0011}-c_{0001} \right) Z_P
    \nonumber\\
    &=& \left( 6 c_{0001}-c_{0011}c_{0111} \right) \left( \frac{1}{2} \kappa_0^2-\frac{1}{2}\kappa_0 \right)+\left( 6c_{0011}-c_{0111}^2 \right)\kappa_0\kappa_1
    \nonumber\\
    &&+ (36c_{0011}-6c_{0111}^2) \frac{\kappa_0}{2}+\frac{1}{2}(15c_{0111}-12c_{01}-\frac{15}{2}c_{0111})
    \nonumber\\
    &\equiv& -N_5(n)Z_{\widetilde{S_{00}}}-N_4^{(1)}(n)Z_{\widetilde{S_{01}}}-\frac{N_4(n)}{2}Z_{\widetilde{C_0}}
\end{eqnarray}
Notice that
\begin{eqnarray}
    c_{01}&=&\frac{5}{8}A_2(n)-\frac{1}{24}N_4^{(1)} \equiv \frac{5}{8}c_{0111}-\frac{1}{24}N_4^{(1)}
\end{eqnarray}
thus the constant term of $-N_4^{(1)}Z_{\widetilde{S_{01}}}$ is indeed given by
\begin{eqnarray}
    \frac{1}{2}(15c_{0111}-12c_{01}-\frac{15}{2}c_{0111})&\equiv&\frac{1}{4}N_4^{(1)}.
\end{eqnarray}
In summary, we find the agreement in monodromy matrices as
\begin{eqnarray}
    M=T\cdot L_{K_Y}\equiv (T_{\calO_X} L_1)^6 L_1^{-6}.
\end{eqnarray}

By the same argument for the "Five Guys" in CY3, we expect that such a decomposition in the theorem 5.1 of \cite{Lin:2026icu} of higher nested torus link also exists in the following CY4 hypersurfaces in the product of projective spaces:
\begin{eqnarray}
    X^\flat&=&[\bbP^5|6],\quad \left[ \begin{array}{c|c}
        \bbP^4 &  5
         \\
        \bbP^1 &  2
    \end{array} \right],\quad \left[ \begin{array}{c|c}
        \bbP^3 &  4
         \\
        \bbP^2 &  3
    \end{array} \right],\quad
    \left[ \begin{array}{c|c}
        \bbP^3 &  4
         \\
        (\bbP^1)^2 &  2
    \end{array} \right],
    \nonumber\\
    &&
    \left[ \begin{array}{c|c}
        (\bbP^2)^2 &  3
         \\
        \bbP^1 &  2
    \end{array} \right],\quad
    \left[ \begin{array}{c|c}
        \bbP^2 &  3
         \\
        (\bbP^1)^3 &  2
    \end{array} \right],\quad
    \left[(\bbP^1)^5|2\right].
\end{eqnarray}

\subsection{$G(2,5)[4,1]$}

\begin{eqnarray}
    X=\left[
\begin{array}{c|ccccc}
\bbP^{m-1} & 1 & \cdots & 1 & 1 & 0\\
G(2,5) & n_1 & \cdots & n_{m-1} & n_m & 1
\end{array}
\right],
\qquad
\sum_{i=1}^{m}n_I=4 .
\end{eqnarray}

Put
\[
A_k(n)=a_kH^k,
\qquad
a_k=e_k(n_1,\ldots,n_m),
\qquad
a_1=4.
\]

The period basis is
\begin{eqnarray}
    \vec\Pi=
\left(
Z_X,
Z_{D_0},
Z_{D_1},
Z_{S_{00}},
Z_{S_{01}},
Z_{S_{11}},
Z_{S_1},
Z_{\widetilde{C_0}},
Z_{\widetilde{C_1}},
Z_P
\right).
\end{eqnarray}

For \(m=2\), the \(S_{00}\)-row and column are omitted.

The relevant intersection numbers are
\[
c_{1111}=20,
\qquad
c_{0111}=5a_2,
\qquad
c_{0011}=5a_3,
\qquad
c_{0001}=5a_4,
\qquad
c_{0000}=0,
\]
and
\[
c^{g}_{111}=8,
\qquad
c^{g}_{011}=2a_2,
\qquad
c^{g}_{001}=2a_3,
\qquad
\int_X\widetilde H^{\,2}=4.
\]

Define
\[
\ell_0=
\frac{37a_3+20a_4-5a_2a_3}{24},
\]
\[
s_{00}=2\ell_0
=
\frac{37a_3+20a_4-5a_2a_3}{12},
\]
and
\[
s_{01}
=
\frac{-5a_2^2+47a_2+35a_3+10a_4}{12}.
\]

The spherical twist by \(\mathcal O_X\) is
\[
T_{\mathcal O_X}
=
\begin{pmatrix}
-1& l_0&7&-s_{00}&-s_{01}&-24&-6&0&0&-1\\
0&1&0&0&0&0&0&0&0&0\\
0&0&1&0&0&0&0&0&0&0\\
0&0&0&1&0&0&0&0&0&0\\
0&0&0&0&1&0&0&0&0&0\\
0&0&0&0&0&1&0&0&0&0\\
0&0&0&0&0&0&1&0&0&0\\
0&0&0&0&0&0&0&1&0&0\\
0&0&0&0&0&0&0&0&1&0\\
0&0&0&0&0&0&0&0&0&1
\end{pmatrix}.
\]

The second spherical object $\calS_X$ is given by
\begin{eqnarray}
    Z_{\calS_X}=(2,0,-1,0,0,0,-1,0,0,0).
\end{eqnarray}
In our basis, $Z_{\calS_X}$ looks very simple due to the defining sequence of $\calO_{S_1}$:
\begin{eqnarray}
    0\rightarrow \calO_X(-D_1)\rightarrow \calS_X \rightarrow \calO_X \rightarrow \calO_{S_1}\rightarrow 0,
\end{eqnarray}
but $Z_{\calS_X}$ computes the same period in \cite{Gerhardus:2016iot} up to sign conventions.

Introduce
\[
t_0=
\frac{
-5a_2^2+5a_2a_3+35a_2-20a_3-10a_4
}{12},
\]
\[
t_{00}
=
\frac{
-5a_2a_3+40a_3-10a_4
}{6},
\]
and
\[
t_{01}
=
\frac{
-5a_2^2+35a_2+20a_3+10a_4
}{6}.
\]

Then
\[
\left(\chi(S_X,B_i)\right)
=
\left(
5,\,
t_0,\,
0,\,
t_{00},\,
t_{01},\,
30,\,
8,\,
0,\,
1,\,
2
\right),
\]
and
\[
T_{S_X}
=
\begin{pmatrix}
-9&-2t_0&0&-2t_{00}&-2t_{01}&-60&-16&0&-2&-4\\
0&1&0&0&0&0&0&0&0&0\\
5&t_0&1&t_{00}&t_{01}&30&8&0&1&2\\
0&0&0&1&0&0&0&0&0&0\\
0&0&0&0&1&0&0&0&0&0\\
0&0&0&0&0&1&0&0&0&0\\
5&t_0&0&t_{00}&t_{01}&30&9&0&1&2\\
0&0&0&0&0&0&0&1&0&0\\
0&0&0&0&0&0&0&0&1&0\\
0&0&0&0&0&0&0&0&0&1
\end{pmatrix}.
\]

The large-volume monodromy
\begin{eqnarray}
    L=
\begin{pmatrix}
1&0&0&0&0&0&0&0&0&0
\\
0&1&0&0&0&0&0&0&0&0
\\
1&0&1&0&0&0&0&0&0&0
\\
0&0&0&1&0&0&0&0&0&0
\\
0&1&0&0&1&0&0&0&0&0
\\
1&0&1&0&0&1&0&0&0&0
\\
0&0&0&0&0&0&1&0&0&0
\\
c_{0111} & c_{0011} & c_{0111}
&
c_{0001}
&
c_{0011}
&
c_{0111}
&
c^{g}_{110}
&
1&0&0
\\
20
&
c_{0111}
&
20
&
c_{0011}
&
c_{0111}
&
20
&
8
&
0&1&0
\\
-10
&
\dfrac{c_{0011}}{2}
&
-10
&
\dfrac{c_{0011}}{2}-c_{0001}
&
-\dfrac{c_{0011}}{2}
&
-10
&
0
&
0&1&1
\end{pmatrix}
\end{eqnarray}

Set
\[
G=T_{S_X}T_{\mathcal O_X}L.
\]

Define
\[
N_5
=
\int_Y\left(A_2A_3-A_1A_4\right)
=
5(a_2a_3-4a_4),
\]
\[
N_4^{(1)}
=
\int_YH\left(A_2^2-A_1A_3\right)
=
5(a_2^2-4a_3),
\]
and
\[
N_4
=
\int_{X^\flat}\left(A_2^2-A_1A_3\right)
=
20(a_2^2-4a_3)
=
4N_4^{(1)}.
\]

For convenience, put
\[
A=
-\frac52
\left(
6a_2^2-a_2a_3-20a_3
\right),
\]
\[
B=-10(a_2-a_3),
\]
and
\[
C=
\frac53
\left(
a_2^2+5a_2-7a_3-2a_4
\right).
\]

Direct multiplication gives
\[
G^4=
\begin{pmatrix}
1&0&0&0&0&0&0&0&0&0\\
0&1&0&0&0&0&0&0&0&0\\
0&0&1&0&0&0&0&0&0&0\\
0&0&0&1&0&0&0&0&0&0\\
0&4&0&0&1&0&0&0&0&0\\
0&-a_2&0&0&0&1&0&0&0&0\\
0&0&0&0&0&0&1&0&0&0\\
0&A&0&-N_5&-N_4^{(1)}&0&0&1&0&0\\
0&B&0&0&0&0&0&0&1&0\\
0&C&0&0&0&0&0&0&0&1
\end{pmatrix}.
\]

Equivalently, the only nontrivial actions are
\[
\begin{aligned}
Z_{D_0}\cdot G^4
={}&
Z_{D_0}
+4Z_{S_{01}}
-a_2Z_{S_{11}}
+A\,Z_{C^f_0}
+B\,Z_{C^f_1}
+C\,Z_P,
\\[2mm]
Z_{S_{00}}\cdot G^4
={}&
Z_{S_{00}}
-N_5Z_{C^f_0},
\\[2mm]
Z_{S_{01}}\cdot G^4
={}&
Z_{S_{01}}
-N_4^{(1)}Z_{C^f_0}.
\end{aligned}
\]

Using
\[
Z_{\widetilde S_{00}}
=
\frac{\kappa_0^2}{2}
-\frac{\kappa_0}{2},
\qquad
Z_{\widetilde S_{01}}
=
\kappa_0\kappa_1-\frac16,
\]
the first transformation can be rewritten as
\[
\begin{aligned}
&
4Z_{S_{01}}
-a_2Z_{S_{11}}
+A\,Z_{C^f_0}
+B\,Z_{C^f_1}
+C\,Z_P
\\
&\hspace{2cm}
=
-N_5Z_{\widetilde S_{00}}
-N_4^{(1)}Z_{\widetilde S_{01}}
-\frac{N_4}{2}Z_{C^f_0}.
\end{aligned}
\]

Therefore
\[
Z_{D_0}\cdot G^4
=
Z_{D_0}
-N_5Z_{\widetilde S_{00}}
-N_4^{(1)}Z_{\widetilde S_{01}}
-\frac{N_4}{2}Z_{C^f_0},
\]
\[
Z_{S_{00}}\cdot G^4
=
Z_{S_{00}}-N_5Z_{C^f_0},
\]
and
\[
Z_{S_{01}}\cdot G^4
=
Z_{S_{01}}-N_4^{(1)}Z_{C^f_0}.
\]

Hence, we find that
\begin{eqnarray}
    (T_{\calS_X}T_{\calO_X}L)^4\equiv T.
\end{eqnarray}

Finally, we show the splitting of degree one hypersurface by tautological bundle which has anomalous multiplicity. The splitting configuration is (the $\yng(1)$ denotes the splitting of $\calO(1)$ by rank two dual tautological bundle on $G(2,5)$)
\begin{eqnarray}
    X=\left[ \begin{array}{c|cc}
        \bbP^1 &  1 & 0
         \\
        G(2,5) & \yng(1) & 4
    \end{array} \right].
\end{eqnarray}
The basis is the same as above, but without the class $\calO_{S_{00}}$ that vanishes for $\bbP^1$. The corresponding matrices are
{
\tiny
\begin{eqnarray}
    T_{\calO_X} &=& \left(
\begin{array}{ccccccccc}
 -1 & 0 & 7 & -6 & -24 & -6 & 0 & 0 & -1 \\
 0 & 1 & 0 & 0 & 0 & 0 & 0 & 0 & 0 \\
 0 & 0 & 1 & 0 & 0 & 0 & 0 & 0 & 0 \\
 0 & 0 & 0 & 1 & 0 & 0 & 0 & 0 & 0 \\
 0 & 0 & 0 & 0 & 1 & 0 & 0 & 0 & 0 \\
 0 & 0 & 0 & 0 & 0 & 1 & 0 & 0 & 0 \\
 0 & 0 & 0 & 0 & 0 & 0 & 1 & 0 & 0 \\
 0 & 0 & 0 & 0 & 0 & 0 & 0 & 1 & 0 \\
 0 & 0 & 0 & 0 & 0 & 0 & 0 & 0 & 1 \\
\end{array}
\right),\qquad T_{\calS_X} = \left(
\begin{array}{ccccccccc}
 -9 & -8 & 0 & -16 & -60 & -16 & 0 & -2 & -4 \\
 0 & 1 & 0 & 0 & 0 & 0 & 0 & 0 & 0 \\
 5 & 4 & 1 & 8 & 30 & 8 & 0 & 1 & 2 \\
 0 & 0 & 0 & 1 & 0 & 0 & 0 & 0 & 0 \\
 0 & 0 & 0 & 0 & 1 & 0 & 0 & 0 & 0 \\
 5 & 4 & 0 & 8 & 30 & 9 & 0 & 1 & 2 \\
 0 & 0 & 0 & 0 & 0 & 0 & 1 & 0 & 0 \\
 0 & 0 & 0 & 0 & 0 & 0 & 0 & 1 & 0 \\
 0 & 0 & 0 & 0 & 0 & 0 & 0 & 0 & 1 \\
\end{array}
\right),
\nonumber\\
L &=& \left(
\begin{array}{ccccccccc}
 1 & 0 & 0 & 0 & 0 & 0 & 0 & 0 & 0 \\
 0 & 1 & 0 & 0 & 0 & 0 & 0 & 0 & 0 \\
 1 & 0 & 1 & 0 & 0 & 0 & 0 & 0 & 0 \\
 0 & 1 & 0 & 1 & 0 & 0 & 0 & 0 & 0 \\
 1 & 0 & 1 & 0 & 1 & 0 & 0 & 0 & 0 \\
 0 & 0 & 0 & 0 & 0 & 1 & 0 & 0 & 0 \\
 8 & 0 & 8 & 0 & 8 & 4 & 1 & 0 & 0 \\
 20 & 8 & 20 & 8 & 20 & 8 & 0 & 1 & 0 \\
 -10 & 0 & -10 & 0 & -10 & 0 & 0 & 1 & 1 \\
\end{array}
\right).
\end{eqnarray}
}
Then it is direct to verify
\begin{eqnarray}
    (T_{\calS_X}T_{\calO_X}L)^4=T^4
\end{eqnarray}

\appendix

\bibliographystyle{fullsort}
\bibliography{biblio.bib}

\end{document}